\documentclass[12pt]{article} 
\pdfoutput=1
\usepackage{amssymb,graphicx}
\usepackage{epstopdf}
\usepackage{amsmath,amsfonts}
\usepackage{epsfig} 
\usepackage{graphicx,graphics}
\usepackage[colorlinks=true, urlcolor=blue, linkcolor=blue]{hyperref}

\usepackage[dvipsnames]{xcolor}

\begin{document}

\title{\bf Biased domain walls: faster annihilation,\\ weaker gravitational waves}
\author{E.~Babichev$^a$, I.~Dankovsky$^{b,c}$, D.~Gorbunov$^{c, d}$, S.~Ramazanov$^{e}$, A.~Vikman$^f$\\
\small{\em $^a$Universit\'e Paris-Saclay, CNRS/IN2P3, IJCLab, 91405 Orsay, France}\\
\small{\em $^b$Faculty of Physics, MSU, 119991 Moscow, Russia}\\
\small{\em $^c$Institute for Nuclear Research of the Russian Academy of Sciences, 117312 Moscow, Russia}\\
\small{\em $^d$Moscow Institute of Physics and Technology, 141700 Dolgoprudny, Russia}\\
\small{\em  $^e$Institute for Theoretical and Mathematical Physics, MSU, 119991 Moscow, Russia}\\
\small{\em $^f$CEICO, Institute of Physics of the Czech Academy of Sciences (FZU),}\\ 
\small{\em Na Slovance 1999/2, 182 00 Prague 8, Czechia}
}
 
 \date{}

{\let\newpage\relax\maketitle}

\begin{abstract}
 We study the evolution of domain wall networks and their phenomenological implications in a model of a real scalar $\chi$, where a $Z_2$-symmetry is slightly broken by a potential bias $V_{bias}$. It is demonstrated that the latter triggers domain wall annihilation considerably earlier than previously thought. Namely, we observe that the scaling relation $t_{ann} \propto 1/V^{2/3}_{bias}$ for the annihilation time $t_{ann}$ fits to the simulation data better than a commonly assumed $t_{ann} \propto 1/V_{bias}$. As a result, the energy density of gravitational waves produced by the network of biased domain walls, for a given tiny $V_{bias}$, is suppressed compared to naive expectations. The spectral shape of gravitational waves is similar to that resulting from unbiased domain walls, but with more power in the close-to-maximum ultraviolet part. 
 In the far ultraviolet region, the spectrum of gravitational waves becomes nearly flat; such a plateau has been recognised earlier in the case of unbiased walls. In our investigation we mainly focus on the symmetry breaking potential $V_{breaking} \propto \chi^3$, and argue that no significant modifications of the domain walls evolution take place if one includes higher powers of $\chi$.  
\end{abstract}

\section{Introduction}
\label{sec:intro}

 Domain walls (DWs) generically arise in extensions of the Standard Model exhibiting spontaneous breaking of discrete symmetries~\cite{Zeldovich:1974uw}. 
 In particular, that situation is typical in scenarios involving a real scalar field having the potential with non-trivial (almost) degenerate minima. For simplicity, we 
  consider the case with $Z_2$-symmetry and two minima. 
 One and sometimes the only possible way of tracing DWs in this and similar scenarios is through their gravitational radiation. 
An intriguing possibility for these gravitational waves (GWs) is that they may have a sufficiently large amplitude to explain
the signal recently found in the pulsar timing array (PTA) data~\cite{NANOGrav:2023gor, NANOGrav:2023hvm, EPTA:2023fyk, EPTA:2023xxk, Xu:2023wog, Reardon:2023gzh}. This is due to the fact that the DW network itself can have a large energy density, which grows relative to that of radiation in the expanding Universe. However, such a growth leads to the unacceptable dominance of 
DWs in the Universe. One of the most common solutions of the problem involves a slight explicit breaking of discrete symmetry by introducing bias in the scalar potential  minima~\cite{Zeldovich:1974uw, Vilenkin:1981zs, Gelmini:1988sf}. The resulting vacuum pressure exerted on the wall eventually leads to the network collapse (for alternative solutions of the DW problem see Refs.~\cite{Vilenkin:1981zs, Lazarides:1982tw, Dvali:1995cc, Coulson:1995nv, Blasi:2022woz, Ramazanov:2021eya, Babichev:2021uvl}). The main goal of this work is to study the impact of the potential bias on DW evolution and GW emission.

It is often considered that the slight symmetry breaking is irrelevant for most of DW evolution, and the network collapse happens almost instantly at a certain cosmological time $t_{ann}$. The latter is commonly inferred from the equality between the energy density of DWs assumed to be in the scaling regime and the constant potential bias $V_{bias}$\footnote{The case of time-dependent potential bias has been considered in Refs.~\cite{Kitajima, Cyr:2025nzf, Notari:2025kqq}, and we do not discuss it here.}, i.e., the difference of potential energy densities of two vacuums. This leads to the following dependence of the annihilation time on the potential bias: $t_{ann} \propto 1/V_{bias}$~\cite{Vilenkin:1981zs}, 
see Sec.~\ref{sec:th} for more details. Furthermore, one usually supposes that the potential bias has a little impact on the shape of GW power spectrum: before the annihilation, at times $t<t_{ann}$, GWs are produced as if $Z_2$-symmetry were exact, and at later times, $t>t_{ann}$, the GW production is entirely terminated and GWs evolve only due to cosmic expansion. This somewhat simplified picture came under scrutiny with high resolution simulations becoming available~\cite{Kitajima, Cyr:2025nzf, Notari:2025kqq, Kawasaki:2014sqa, Kitajima:2023kzu, Pujolas}\footnote{In the early work on the subject~\cite{Larsson:1996sp}, the issue with limited resolution was circumvented by assuming the wall width growing with time. 
This technique developed in Ref.~\cite{Press:1989yh} allowed for long simulation times. We do not use it in the present work.}.

Our numerical simulations performed using the public code CosmoLattice~\cite{Figueroa1, Figueroa2} reveal 
the behavior
\begin{equation}
\label{tann}
t_{ann} \propto 1/V^{2/3}_{bias} \; , 
\end{equation}
in contrast to aforementioned $t_{ann} \propto 1/V_{bias}$. Note that values of the 
potential bias $V_{bias}$ are typically assumed to be very (almost exponentially) small, and therefore this deviation from naive expectations can have a profound impact on 
GW phenomenology. Namely, for the same $V_{bias}$ one obtains a much earlier decay of the DW network compared to naive estimates, and hence a weaker GW signal, because the energy density of DWs has less time to grow relative to the dominant, e.g., radiation, energy density. Note that the result~\eqref{tann} has been derived for not 
extremely small $V_{bias}$, see Sec.~\ref{sec:dw}. 
This has a consequence that the time span of simulations prior to the DW annihilation is rather limited, and there is no enough time 
for the scaling regime to be properly established. In this situation, the evolution can be sensitive to the choice of initial conditions. We address these shortcomings in Sec.~\ref{sec:dw} by varying initial conditions and by including the data obtained in Ref.~\cite{Pujolas}, where 
higher resolution simulations allowed to probe much smaller values of $V_{bias}$ compared to our case.

The shapes of GW power spectra obtained with and without potential bias~\cite{Hiramatsu:2013qaa, Ferreira:2023jbu, Dankovsky:2024zvs} are similar, but there are notable differences. In both cases, there is a peak at frequency $f_{peak}$ defined by the correlation length of the DW network, which is of the order of the Hubble radius at annihilation. The far infrared (IR) part of the spectrum
is mostly determined by causality considerations rather than particularities of DW evolution with the characteristic slope approaching $\Omega_{gw} \propto f^3$~\cite{Caprini:2009fx, Cai:2019cdl} at $f\ll f_{peak}$. In the close-to-maximum ultraviolet (UV) part of the spectrum, $f\gtrsim f_{peak}$, we observe a significantly softer decrease of the spectrum compared to the unbiased case. A similar finding has been made recently in Ref.~\cite{Cyr:2025nzf}. Interestingly, the plateau
in the far UV region, $f\gg f_{peak}$, already present in the unbiased case~\cite{Dankovsky:2024zvs}, also occurs for biased DWs. The spectrum ends with the characteristic exponential falloff at frequencies exceeding the inverse DW width, which is also a universal feature.

The paper is organised as follows. In Sec.~\ref{sec:th}, we introduce the model and various notations, and discuss expectations regarding DW evolution and GW emission. We set up the system for numerical simulations in Sec.~\ref{sec:nsetup}. Results of numerical simulations for DWs and GWs are discussed in Secs.~\ref{sec:dw} and~\ref{sec:gw}, respectively. 
We conclude in Sec.~\ref{sec:conclusions}.

\section{Theoretical setup}
\label{sec:th}

We consider the following simplest model of the real scalar field $\chi$, where biased DWs arise:
\begin{equation}
S=\int d^4 x \sqrt{-g} \left[ \frac{1}{2} (\partial_{\mu} \chi)^2 -\frac{1}{4} \cdot \lambda (\chi^2 -v^2)^2 -V_{breaking} \right] \; .
\end{equation}
Here $\lambda$ is the quartic coupling constant and $v$ is the expectation value responsible for spontaneous breaking of $Z_2$-symmetry. The term $V_{breaking}$ explicitly breaks the $Z_2$-symmetry. We choose it to be of the cubic form, 
\begin{equation}
\label{breaking}
V_{breaking}=\epsilon \chi^3 \; ,
\end{equation}
where $\epsilon$ is a constant of the mass dimension referred to as the bias parameter in what follows. We discuss other choices of the symmetry breaking in Sec.~\ref{sec:dw}. The resulting potential bias is given by 
\begin{equation}
V_{bias}=V_{breaking} (\chi \approx v)-V_{breaking} (\chi \approx -v) \approx 2\epsilon v^3 \; ,
\end{equation}
where we have assumed $V_{bias} \ll \lambda v^4$, so that $\epsilon \ll \lambda v$. We are interested in the field $\chi$ evolution in the spatially flat FLRW Universe described by metric  
\begin{equation}
ds^2 =dt^2 -a^2 (t) d{\bf x}^2 \; ,
\end{equation}
where $a(t)$ is the Universe scale factor. Throughout this work, 
we assume that DWs live in the radiation-dominated Universe with the scale factor growing as $a(t) \propto \sqrt{t}$. Below we often make use of the conformal time $\tau$ defined as $\tau \sim \int dt/a(t)$, so that $a(\tau) \propto \tau$.

Let us first briefly review basics of DW evolution in the limit $V_{bias} \rightarrow 0$, where the exact $Z_2$-symmetry is restored. We assume that the field $\chi$ is initially set to zero cosmologically and starts rolling to its potential minima after the Hubble parameter $H=\dot a/a$ drops enough to fulfill the condition $H \lesssim\sqrt{\lambda} v$, from which point on the DW network comes into existence. Note that we neglect any non-gravitational interactions of $\chi$ with the primordial plasma, and hence its thermal mass is set to zero.  
Soon after its formation, the DW network enters the scaling regime~\cite{Press:1989yh}. Numerical simulations~\cite{Dankovsky:2024zvs} suggest that the scaling starts independently of initial conditions at the time $t_{sc}$, when the DW width $\delta_{wall}$, estimated as 
\begin{equation}
\delta_{wall} \simeq \sqrt{\frac{2}{\lambda}} \cdot \frac{1}{v}\,,
\end{equation}
becomes small compared to the Hubble radius, i.e., $\delta_{wall} \simeq 0.05 H^{-1}_{sc}$. In the scaling regime, the long-range dynamics of the DW network is defined by only one parameter, the expansion rate of the Universe, $H$. In particular, simulations reveal that at any time there is essentially one long DW with a characteristic curvature radius $H^{-1}$ in the Hubble patch. Closed DWs are also present, but their contribution to the overall network is negligible~\cite{Dankovsky:2024zvs}. As a result, the energy density of the network is estimated as
\begin{equation}
\label{energysc}
\rho_{wall} \sim \sigma_{wall} H \; ,
\end{equation}
where $\sigma_{wall}$ is the DW tension:
\begin{equation}
\sigma_{wall}=\frac{2\sqrt{2 \lambda} v^3 }{3} \; .
\end{equation}
The latter is defined as the integral over the DW profile (kink), i.e., $\sigma_{wall}=
\int^{+\infty}_{-\infty} dz' T_{00} (z')$, where $T_{00} (z)$ is the $00$the component of the field $\chi$ stress-energy tensor. As it follows from Eq.~\eqref{energysc}, the energy density of DWs grows fast relative to that of radiation, i.e., $\rho_{wall}/\rho_{rad} \propto a^2$, where $a$ is the Universe scale factor. This leads to the DW problem and motivates the introduction of the potential bias.

From now on we switch to the case of biased DWs, unless otherwise specified. Conventionally, one defines the DW network annihilation time $\tilde{\tau}_{ann}$ (the reason for the tilde notation will become clear shortly) from the balance between the DW energy density and the potential bias, i.e.,
$\rho_{wall} \sim V_{bias}$. One also assumes that the scaling of DWs continues up to the annihilation time, so that $\rho_{wall} (\tilde{\tau}_{ann}) \sim \sigma_{wall} H_{ann}$. This leads to the dependence $\tilde{\tau}_{ann} \propto 1/\sqrt{\epsilon}$ of so defined annihilation time $\tilde{\tau}_{ann}$~\cite{Vilenkin:1981zs}. However, we observe in Sec.~\ref{sec:dw} that departures from the scaling regime start 
significantly earlier, thus compromising the physical interpretation of $\tilde{\tau}_{ann}$. Therefore, we consider an alternative way of defining the DW network collapse time, which will prove to be more relevant. Following Refs.~\cite{Kitajima:2023kzu, Pujolas}, we introduce the false vacuum fraction: 
\begin{equation}
\label{fvf}
{\cal F}_{fv}=\frac{V_{false}}{V} \; ,
\end{equation}
where $V$ and $V_{false}$ denote the total volume (of the simulation box) and the volume filled with the false vacuum, respectively. Initially, 
the false vacuum fraction reads ${\cal F}_{fv}=1/2$, and then decreases with time reflecting decay of the false vacuum. Following the definition of the lifetime of elementary particle, one defines the annihilation time of the DW network as the moment $\tau_{ann}$, when the fraction ${\cal F}_{fv}$ reduces by the factor $e$, i.e., 
\begin{equation}
\label{anndef}
{\cal F}_{fv} (\tau_{ann})\equiv \frac{1}{2e} \; .
\end{equation}
Naively one would expect that $\tau_{ann} \sim \tilde{\tau}_{ann} \propto 1/\sqrt{\epsilon}$. 
However, we will see in Sec.~\ref{sec:dw} that the simulation data favor a different behavior. 
Hereafter we adopt the definition of the DW annihilation time~\eqref{anndef}.

DWs are powerful sources of stochastic GWs. The latter are commonly described by the fractional spectral energy density: 
\begin{equation}
\label{def}
\Omega_{gw} \equiv \frac{1}{\rho_{tot}} \cdot \frac{d\rho_{gw}}{d\ln f} \; ,
\end{equation}
where $d\rho_{gw}/d\ln f$ is the spectral energy density of GWs defined per logarithm of frequency $f$, and $\rho_{tot}=3H^2 M^2_{P}$ is the total energy density of the Universe. We use the reduced Planck mass $M_{P} \approx 2.44 \cdot 10^{18}~\mbox{GeV}$.
In the unbiased case, the peak frequency of GWs generated by the conformal time $\tau$ is estimated as~\cite{Dankovsky:2024zvs} (cf. Ref.~\cite{Hiramatsu:2013qaa})
\begin{equation}
\label{estpeak}
\frac{k_{peak}}{2\pi a (\tau)} \simeq 0.7 H (\tau) \; .
\end{equation}
The corresponding peak energy density of GWs is estimated as~\cite{Dankovsky:2024zvs} 
\begin{equation}
\label{estpeakgw}
\Omega_{gw, peak} (\tau) \simeq \frac{3 \cdot 10^{-3} \cdot \sigma^2_{wall} }{ H^2 (\tau) \cdot M^4_{P}}  \; .
\end{equation}
In the case of biased DWs, production of GWs is terminated at $\tau \sim \tau_{ann}$.
Since only last instants, prior to DW disappearance, matter for GW production, one may roughly estimate the GW characteristics by substituting $\tau \sim \tau_{ann}$ into Eqs.~\eqref{estpeak} and~\eqref{estpeakgw}. However, in practice, the subtle details of DW collapse may significantly affect the GW signal. 
Moreover, as it has been mentioned above, violations of the scaling regime assumed in Eqs.~\eqref{estpeak} and~\eqref{estpeakgw} start 
well before the time $\tau_{ann}$, cf. Sec.~\ref{sec:dw}. This further questions the applicability of Eqs.~\eqref{estpeak} and~\eqref{estpeakgw} in the case 
of biased DWs.


\section{Numerical setup}
\label{sec:nsetup}
Numerical simulations are performed with the public code CosmoLattice~\cite{Figueroa1, Figueroa2}. We make use of lattices 
of different resolution, with $1024^3$ and $2048^3$ lattice sites. The simulation box with the comoving size $L$ mimics radiation dominated Universe and  
contains many causally disconnected patches. Simulations are initiated at the time when $H_i=\sqrt{\lambda}v$. We switch to dimensionless variables and model constants:
\begin{equation}
\chi \rightarrow \frac{\chi}{v}\,, \qquad \tau \rightarrow \sqrt{\lambda} v \tau\,, \qquad 
x_i \rightarrow \sqrt{\lambda} v x_i\,, \qquad \epsilon \rightarrow \frac{\epsilon}{\lambda v} \; .
\end{equation}
Setting the initial scale factor $a_i=1$, one obtains from the condition $H_i =\sqrt{\lambda} v$ that the initial dimensionless 
conformal time is 
$\tau_i=1$. The equation of motion for the field $\chi$ in dimensionless variables takes the form: 
\begin{equation}
\chi'' +\frac{2}{\tau} \chi'-\partial^2_i \chi+\tau^2 \chi (\chi^2-1) +3\epsilon \tau^2 \chi^2 =0 \; .
\end{equation}
In what follows we exploit the initial conditions described by 
\begin{equation}
\label{init}
\langle \chi^2 ({\bf x}) \rangle =\int^{k_{cut}}_{k_{min}} \frac{dk\cdot k}{4\pi^2}\,, \qquad \langle \dot{\chi}^2 ({\bf x}) \rangle =\int^{k_{cut}}_{k_{min}} \frac{dk\cdot k^3}{4\pi^2} \; .
\end{equation}
Here $k_{cut}$ and $k_{min}$ are the momentum 
upper and lower cutoffs, respectively. While $k_{min}$ is fixed by the lattice size, i.e., $k_{min}=2\pi/L$, we keep $k_{cut}$ flexible. In this way one can test whether the evolution of biased DWs is independent of initial conditions, --- the property, which  has been demonstrated in the case of unbiased DWs\,\cite{Dankovsky:2024zvs}. We nominally refer to initial conditions~\eqref{init} as vacuum ones.

Optimally by the end of simulations at the conformal time $\tau_f$, the DW width should not be smaller than the lattice spacing, i.e., 
\begin{equation}
\label{cond1}
\delta_{wall} =\sqrt{\frac{2}{\lambda}} \cdot \frac{1}{v} =\frac{\kappa L_i }{N} \cdot \frac{a(\tau_f)}{a(\tau_i)} \; ,
\end{equation}
where $N=1024$ or $2048$ is the lattice grid number, $\kappa = {\cal O} (1)$ is a constant, $L_i=a(\tau_i) L=L$ is the initial lattice size, so that $L_i/N$ is the initial lattice spacing. We take into account that the lattice spacing grows linearly with the scale factor. We also require that the final Hubble horizon is still smaller than the simulation box, 
i.e., 
\begin{equation}
\label{cond2}
H^{-1} (\tau_f) =\frac{L_i}{2\kappa'} \cdot \frac{a(\tau_f)}{a(\tau_i)} \; ,
\end{equation}
where we have introduced another constant $\kappa' = {\cal O} (1)$. Hence, 
\begin{equation}
H^{-1} (\tau_i) = \frac{1}{\sqrt{\lambda}v} =\frac{\kappa}{\sqrt{2}} \cdot \frac{L_i}{N} \cdot \frac{a(\tau_f)}{a(\tau_i)} \; .
\end{equation}
Then we get for the time duration of simulations (recall that $a\propto \tau$ and $H\propto \tau^{-2}$): 
\begin{equation}
\frac{\tau_f}{\tau_i} =\frac{\sqrt{N}}{2^{1/4} \sqrt{\kappa \kappa'}} \; .
\end{equation}
Consequently, 
\begin{equation}
H^{-1} (\tau_i) =\frac{1}{\sqrt{\lambda}v}=\frac{2^{1/4} \sqrt{\kappa} a(\tau_i) L}{2\sqrt{ \kappa' N}} \; .
\end{equation}
We impose this condition in what follows.

Note that for $\kappa \simeq \kappa' \simeq 1$, we have 
\begin{equation}
\label{finaltime}
\frac{\tau_f}{\tau_i} \simeq  35~(N=2048)\,, \qquad \frac{\tau_f}{\tau_i} \simeq 25~(N=1024)  \; .
\end{equation}
Below we fix these final times, but allow for variations of $\kappa'$ and $\kappa$. For concreteness, we restrict to three possible choices of $\kappa'$ and $\kappa$:
\begin{equation}
\label{zoom}
\kappa'=\pi \kappa~\mbox{(IR)}\,,  \qquad \kappa'=\frac{\pi \kappa}{2} ~\mbox{(middle)}\,, \qquad \kappa'=\frac{\pi \kappa}{6} ~\mbox{(UV)}\; .
\end{equation}
The reason to consider $\kappa'/\kappa>1$ is that DW phenomenology and GW signal in particular are dominated by the Hubble scale physics at the latest stages of the network existence. Therefore, we wish the simulation box size to be considerably larger than the Hubble radius at those times, i.e., we allow the ratio $\kappa'/\kappa$ to be larger than unity. With such a choice, however, one runs the risk that DWs can become too thin relative to the lattice spacing by the end of simulations. As it has been observed in Ref.~\cite{Dankovsky:2024zvs}, this does not invalidate simulations, but leads to the appearance of the artificial features in the UV part of the GW spectrum 
propagating towards IR with time. The ``contamination'' is mitigated by choosing the option with $\kappa'/\kappa <1$ in Eq.~\eqref{zoom}.

While we mainly take $\kappa'=\pi \kappa/2$ in what follows, in order to reconstruct the GW spectrum, one will need to consider all three options in Eq.~\eqref{zoom}. In this way, we can ``zoom in'' the IR, middle, and UV parts of the GW spectrum, and then glue all three parts.  
The validity of this techniques was proven in Ref.~\cite{Dankovsky:2024zvs}, where unbiased DWs have been studied 
by comparing results obtained with lattices of different resolution. 

\section{Numerical results: evolution of domain wall network}
\label{sec:dw}

\begin{figure}[!htb]
\begin{center}
    \includegraphics[width=0.45\textwidth]{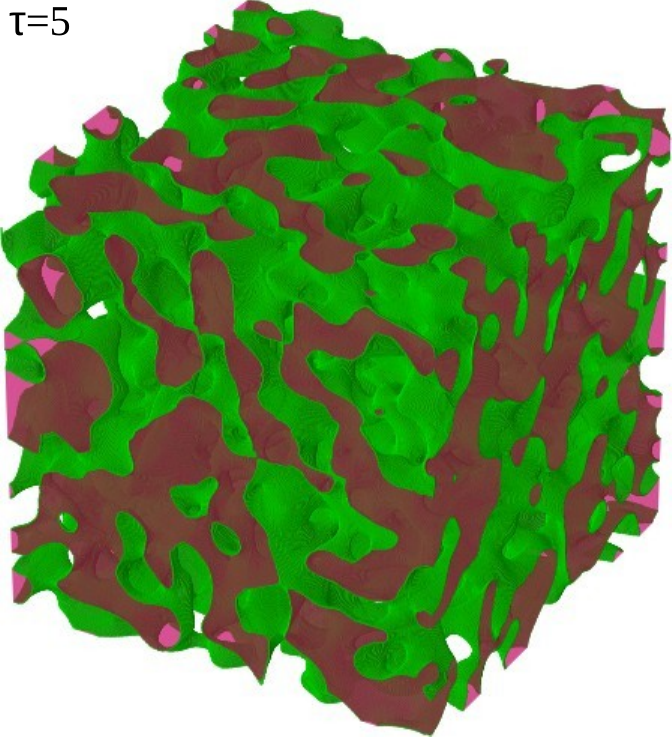} 
    \includegraphics[width=0.45\textwidth]{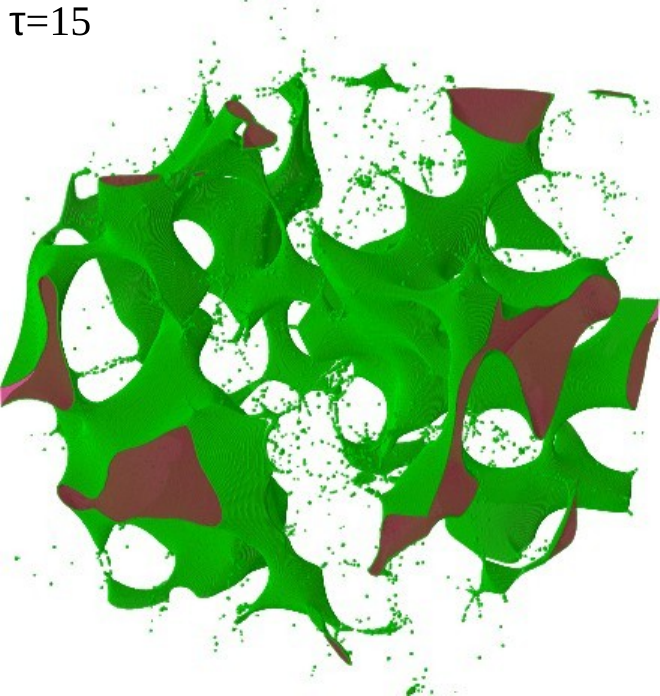} 
    \includegraphics[width=0.45\textwidth]{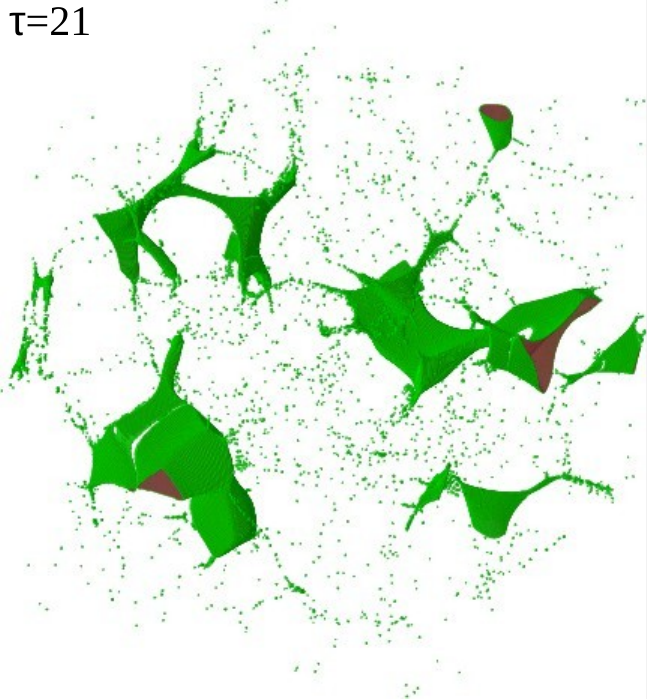} 
 \includegraphics[width=0.45\textwidth]{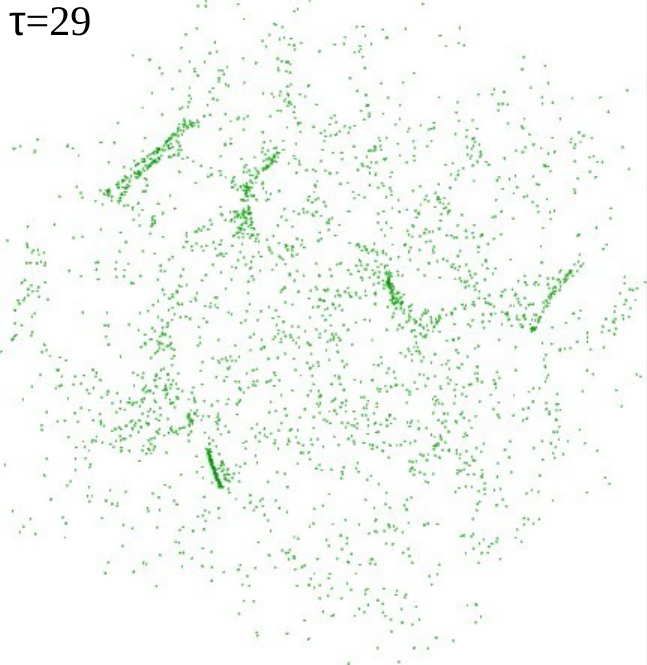}    
\end{center}
    \caption{Snapshots of DW evolution obtained in the case of dimensionless bias parameter $\epsilon=0.01$. DWs are shown with green, while the regions filled with true and false vacuum are shown with white and red, respectively.} \label{snaps}
\end{figure}

\begin{figure}[!htb]
\begin{center}
    \includegraphics[width=0.3\textwidth]{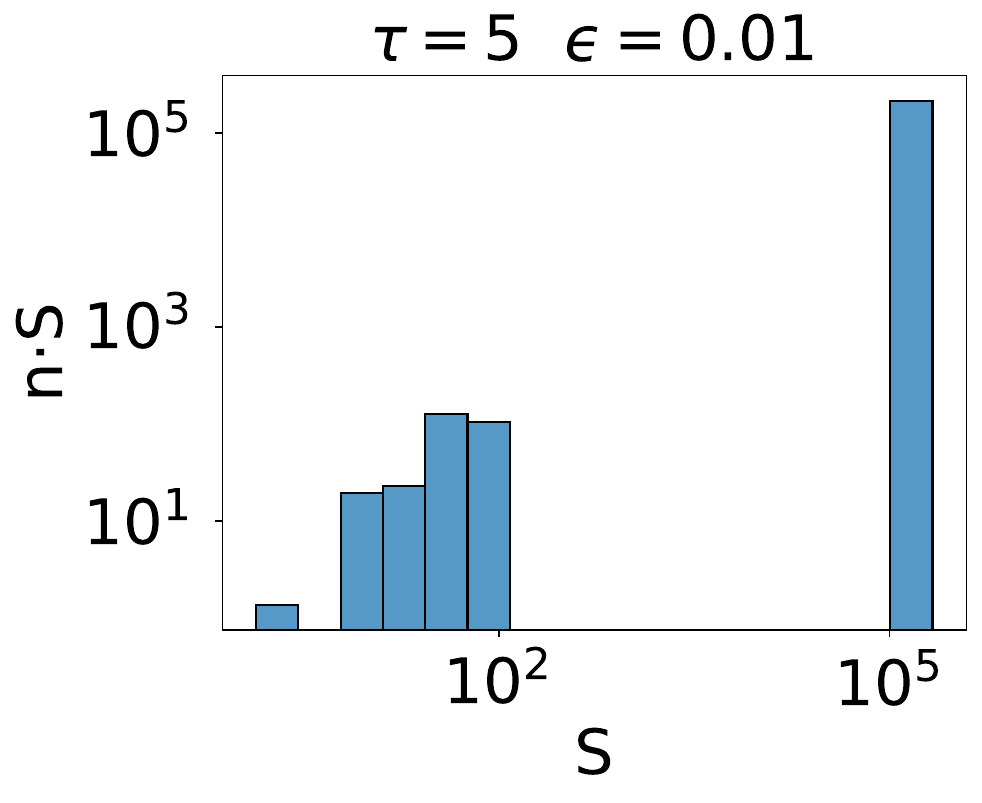} 
    \includegraphics[width=0.3\textwidth]{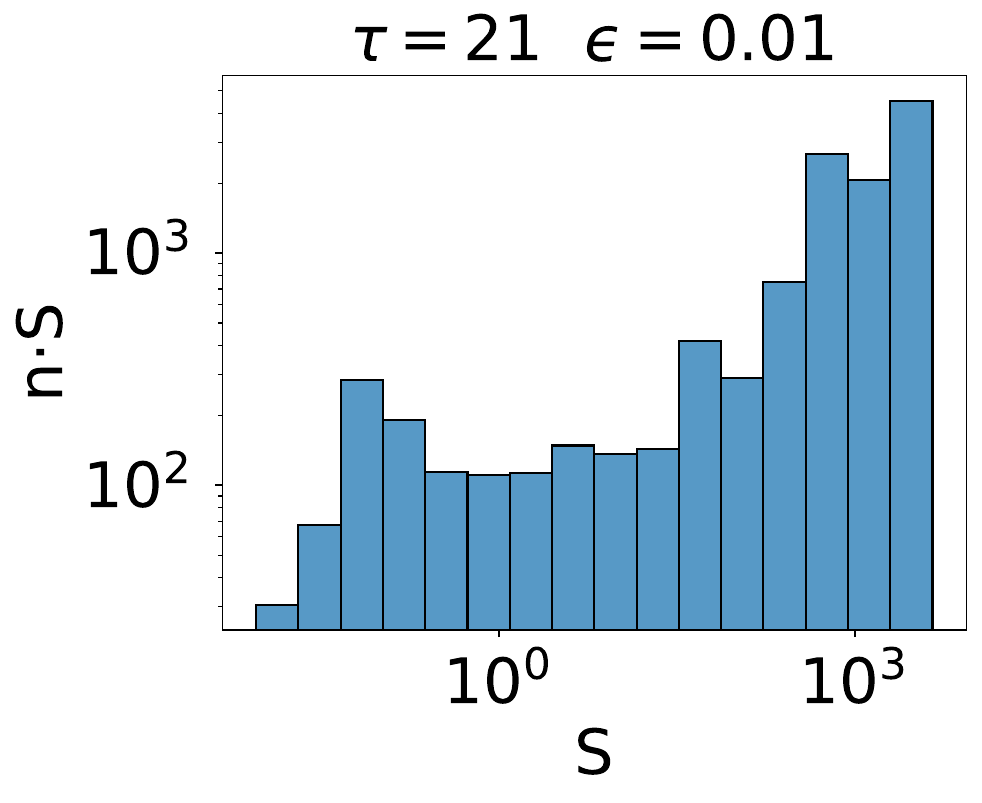} 
    \includegraphics[width=0.3\textwidth]{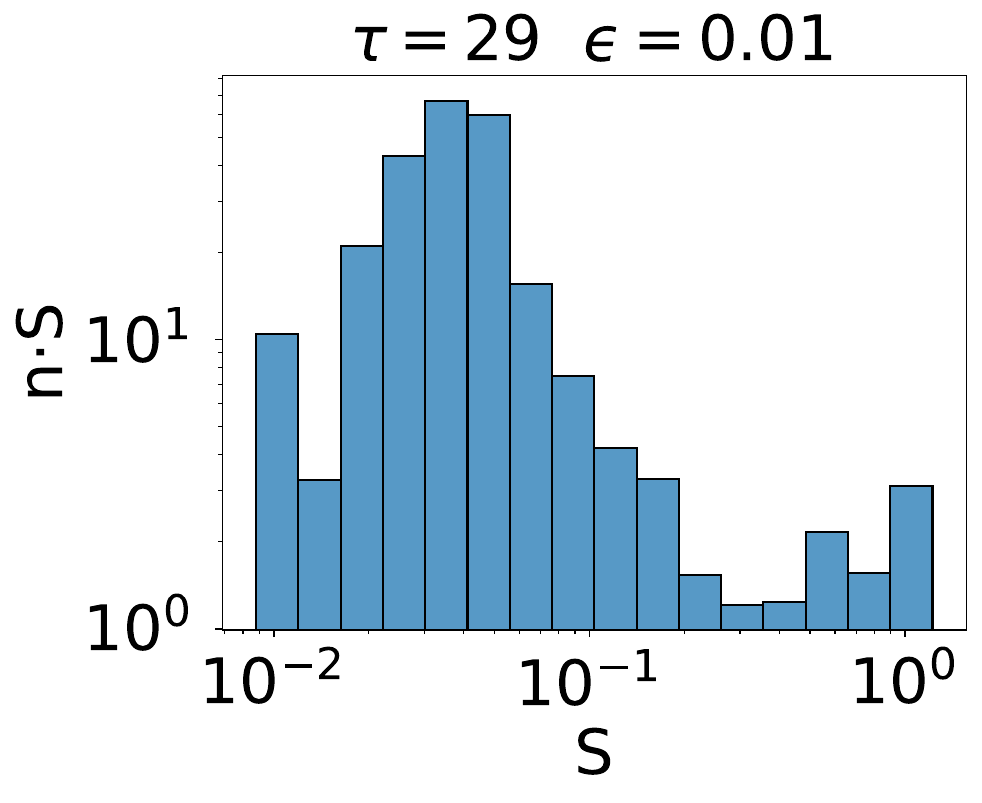} 
        \includegraphics[width=0.3\textwidth]{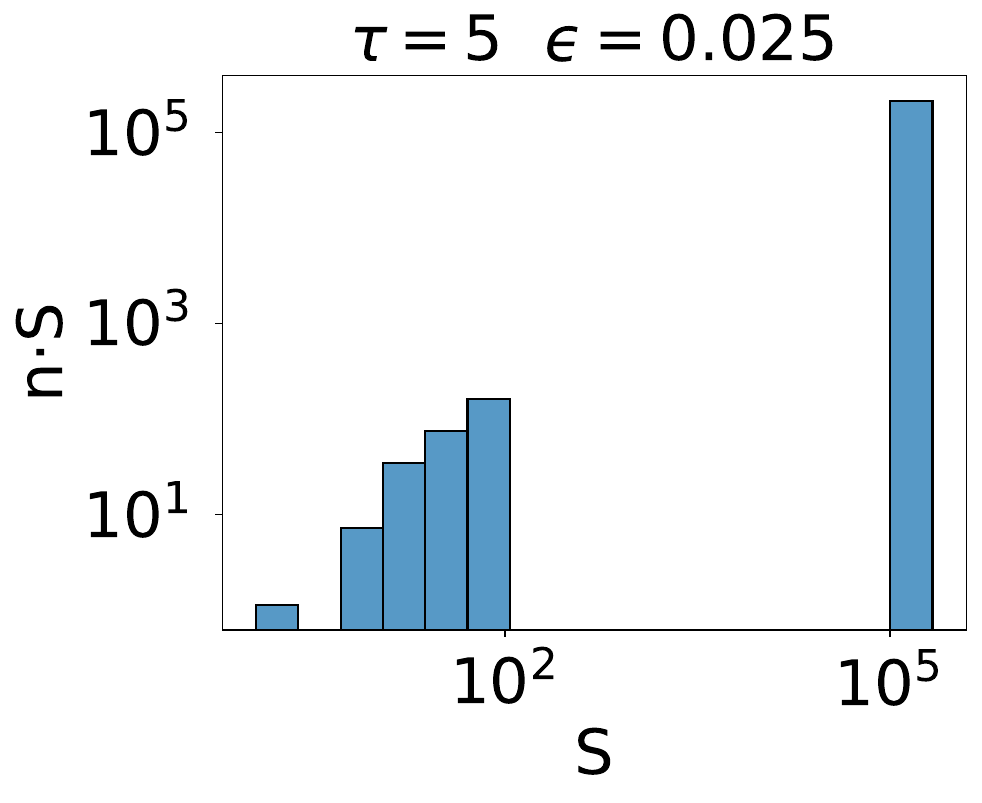} 
    \includegraphics[width=0.3\textwidth]{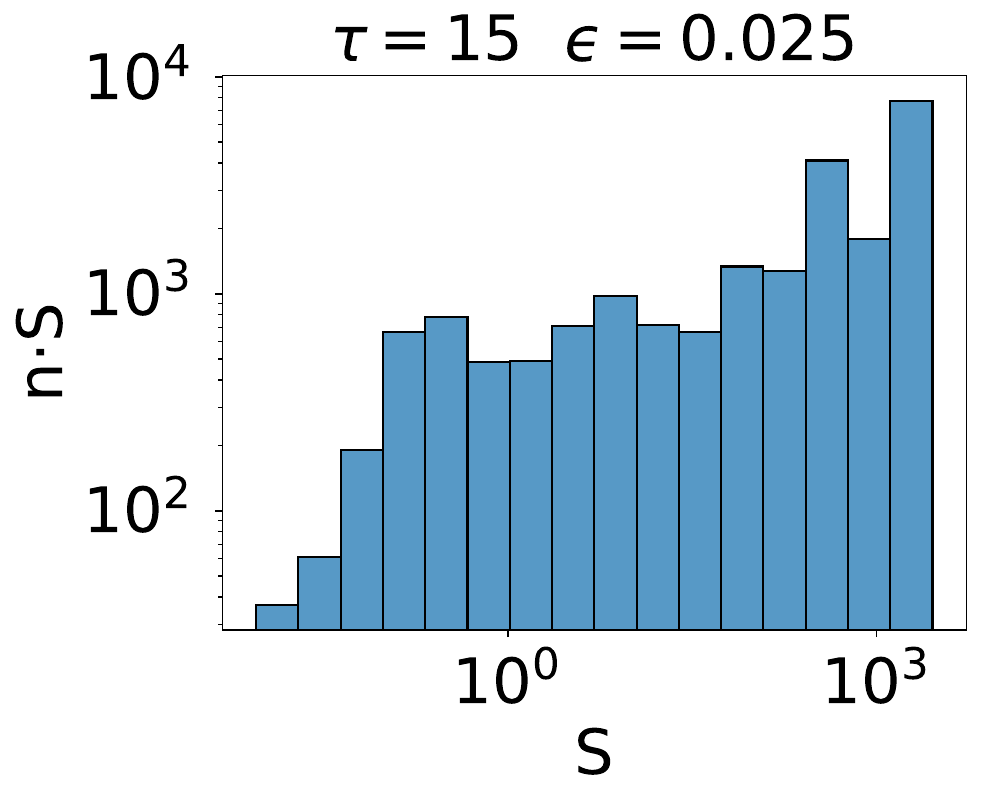} 
        \includegraphics[width=0.3\textwidth]{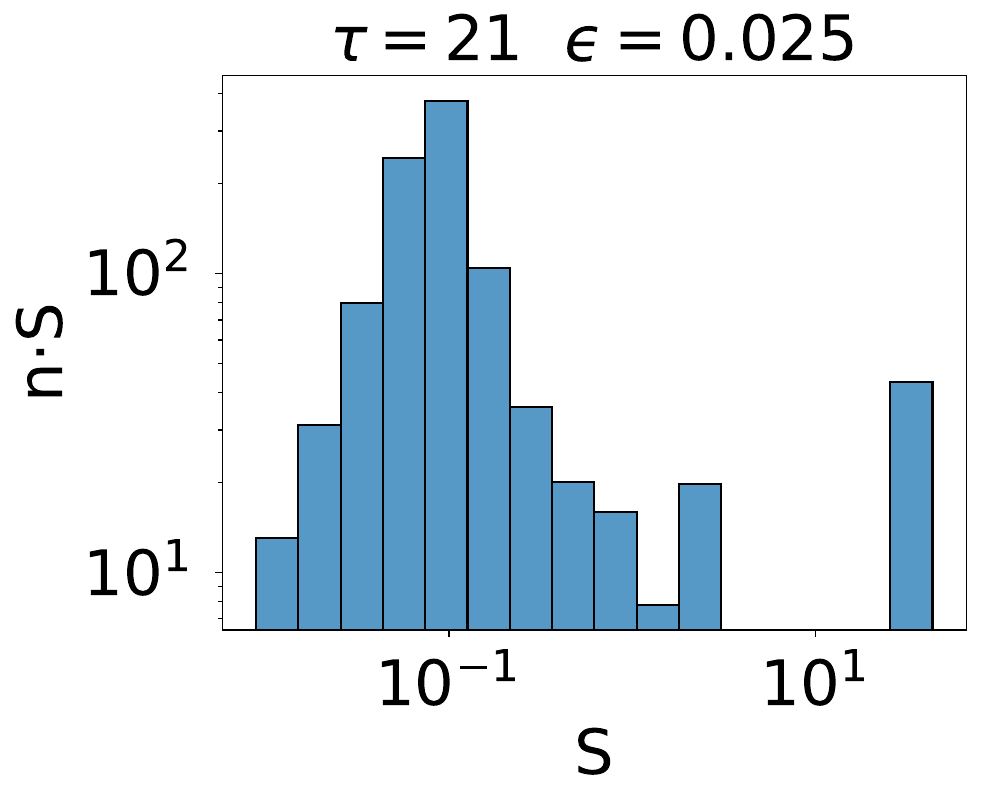} 
    \includegraphics[width=0.3\textwidth]{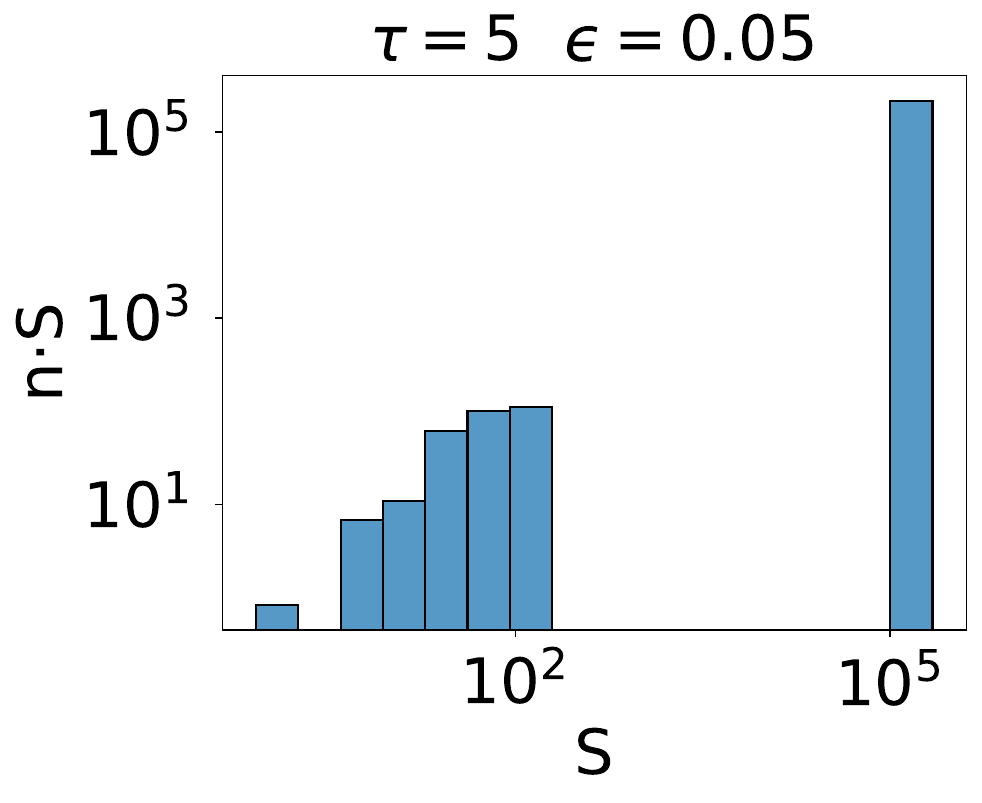} 
    \includegraphics[width=0.3\textwidth]{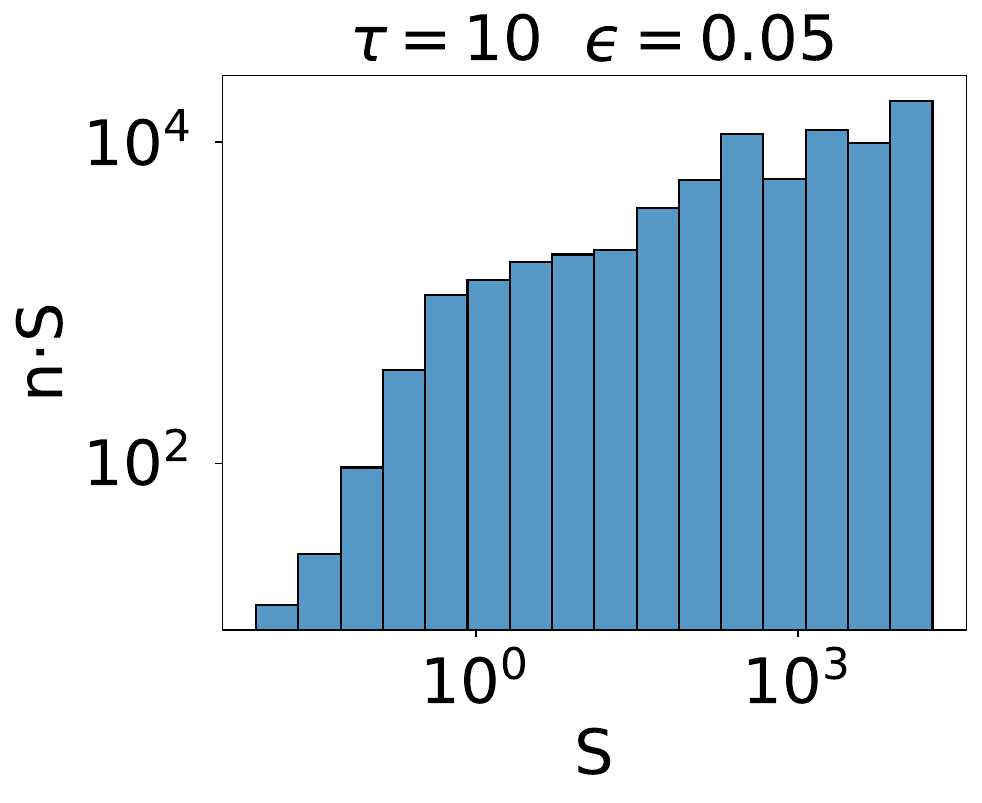} 
    \includegraphics[width=0.3\textwidth]{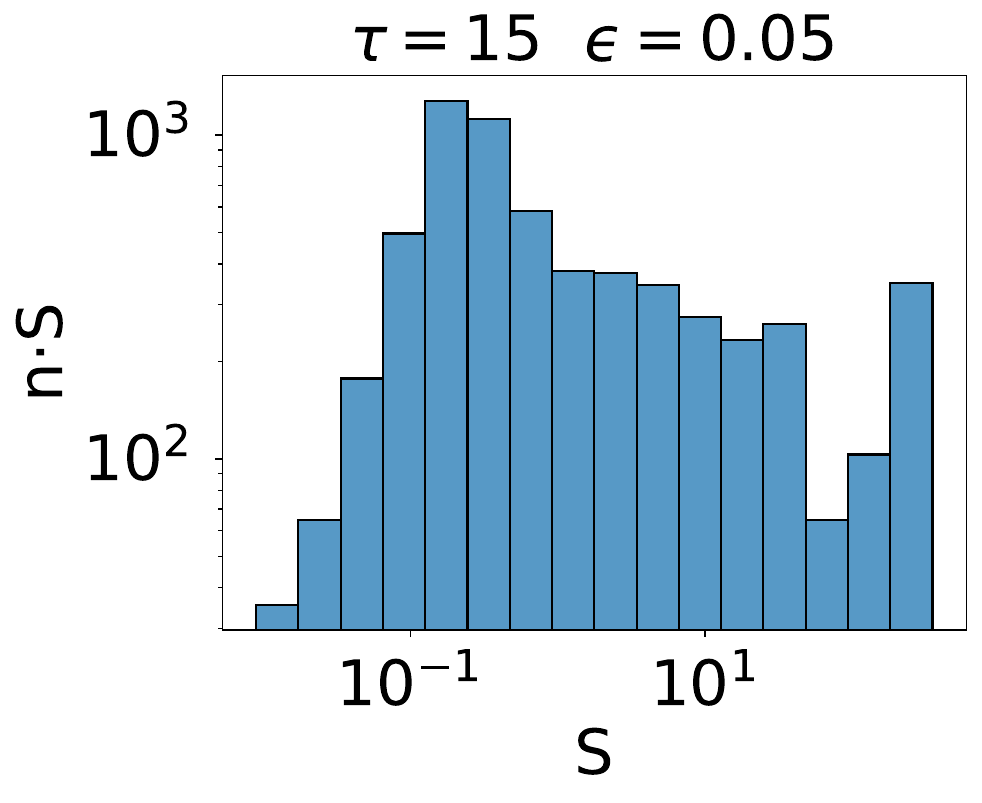} 
\end{center}
    \caption{Histograms showing distribution of the DW network over the area $S$ for three different values of the bias parameter $\epsilon$. Simulations have been carried out on $2048^3$ lattice.} \label{histograms}
\end{figure}

The DW network with biased potential rapidly decays. 
Snapshots of DW evolution obtained in the case of dimensionless bias parameter $\epsilon=0.01$ are shown in Fig.~\ref{snaps}. At initial stages of evolution, there is an equipartition between true and false vacuums separated by a complicated wall network. Later on, larger walls get dissected into smaller walls eventually collapsing and producing scalar radiation. This is illustrated by 
distribution of DWs over areas $S$ depicted with histograms in Fig.~\ref{histograms}. One observes the following stages of DW evolution. Around the time $\tau \sim 5$, there is a long wall separated from smaller closed walls by a gap. This picture is very similar to what one has in the case of unbiased DWs~\cite{Dankovsky:2024zvs}. However, in the case of biased DWs the gap is shrinking with time and one eventually gets the continuous ungapped distribution of DWs over the area. Later on, this DW distribution takes the bell shape. The overall area of DWs is shrinking fast following the rapid collapse of closed walls into particles $\chi$.

\begin{figure}[!htb]
\begin{center}
\includegraphics[width=0.9\textwidth]{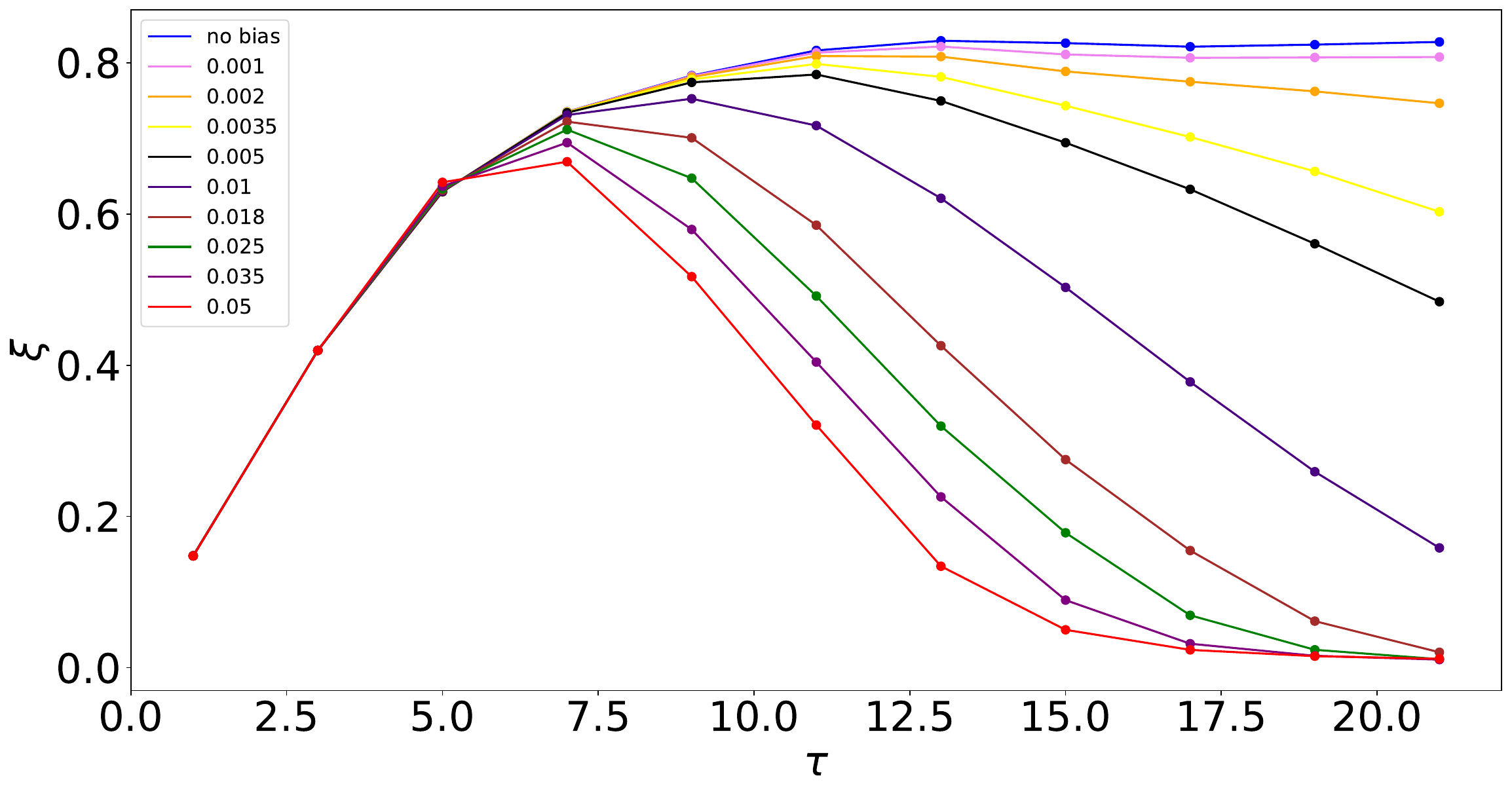} 
\end{center}
    \caption{Scaling parameter $\xi$ defined in Eq.~\eqref{scaling} is shown for different values of the bias parameter $\epsilon$. Simulations have been carried out with $1024^3$ lattice.} \label{fig:scaling}
\end{figure}

In Fig.~\ref{fig:scaling}, DW evolution is described in terms of the scaling (or area) parameter $\xi$ defined as 
\begin{equation}
\label{scaling}
\xi \equiv \frac{S t}{a(t)V} \; ,
\end{equation}
where $S$ is the DW comoving area captured within the comoving volume $V$. The numerical computation of the DW area $S$ is performed using the estimator developed in Ref.~\cite{Press:1989yh}. Note that in the unbiased case the parameter $\xi$ serves to diagnose when the scaling regime is established. Namely, once the scaling is reached, the parameter $\xi$ takes on a constant value. In the case $\epsilon \neq 0$, DW evolution remains approximately unaffected by the potential bias until the time $\tau \sim 5$, which reiterates the above observation made with histograms. Around the time $\tau \sim 5$, different $\xi(\tau)$-curves start to deviate from the case $\epsilon =0$. Overall, the $\xi(\tau)$-curves take bell-like shapes (for vacuum initial conditions), which are more pronounced for larger $\epsilon$. The bell width is growing upon the decrease of the bias parameter.

\begin{figure}[!htb]
\begin{center}
    \includegraphics[width=\textwidth]{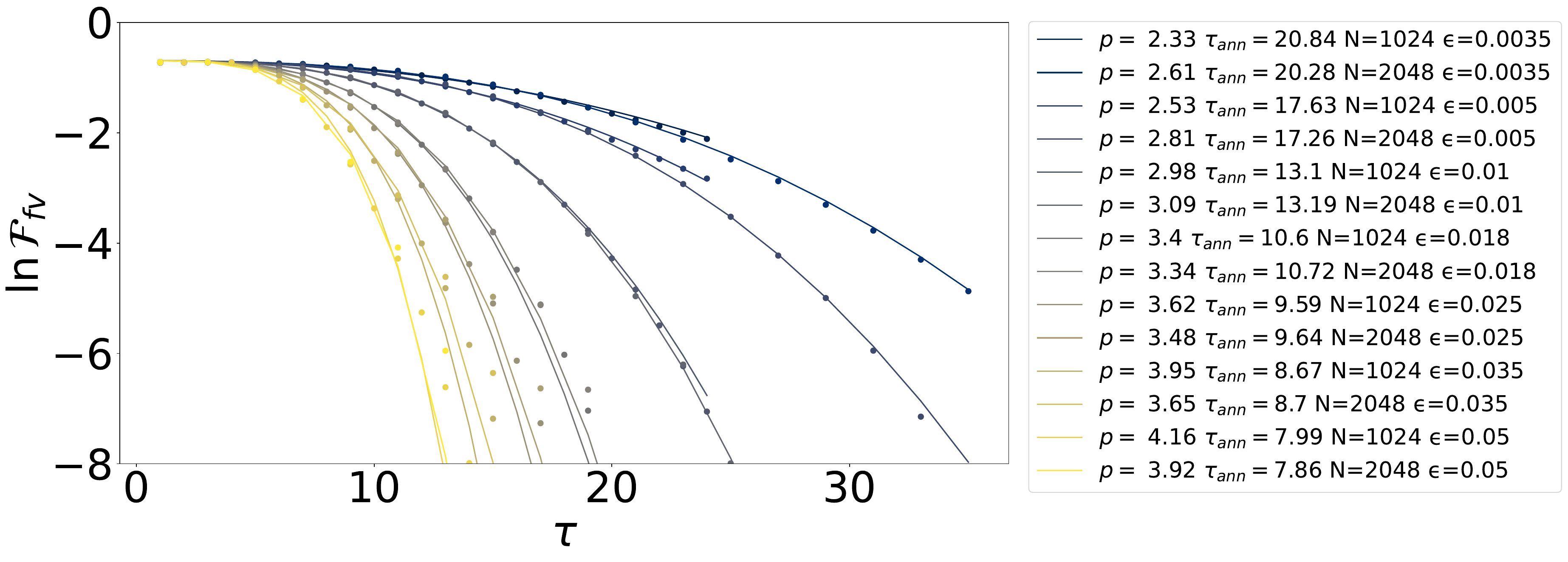} 
        \includegraphics[width=0.492 \textwidth]{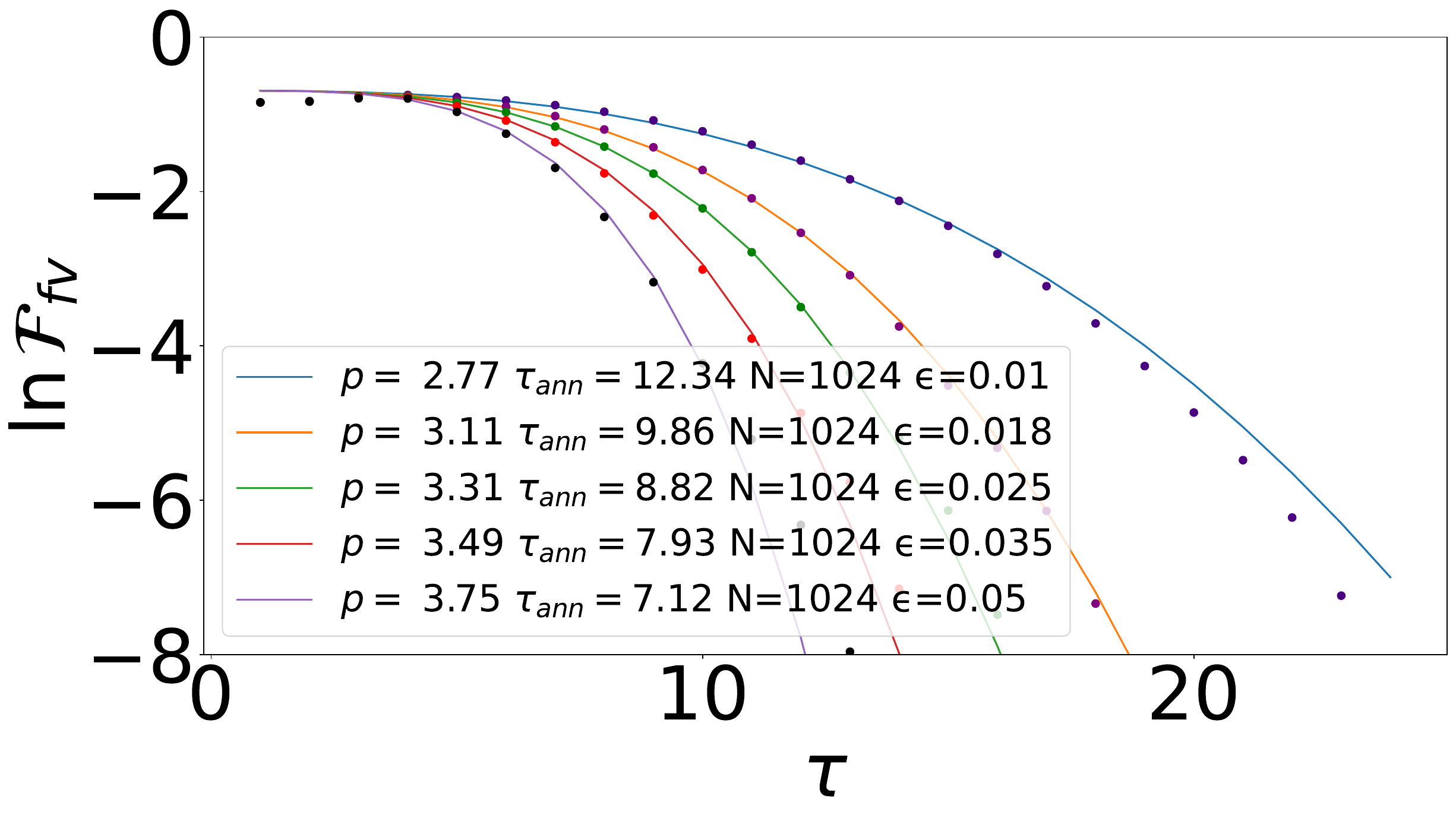} 
  \includegraphics[width=0.492\textwidth]{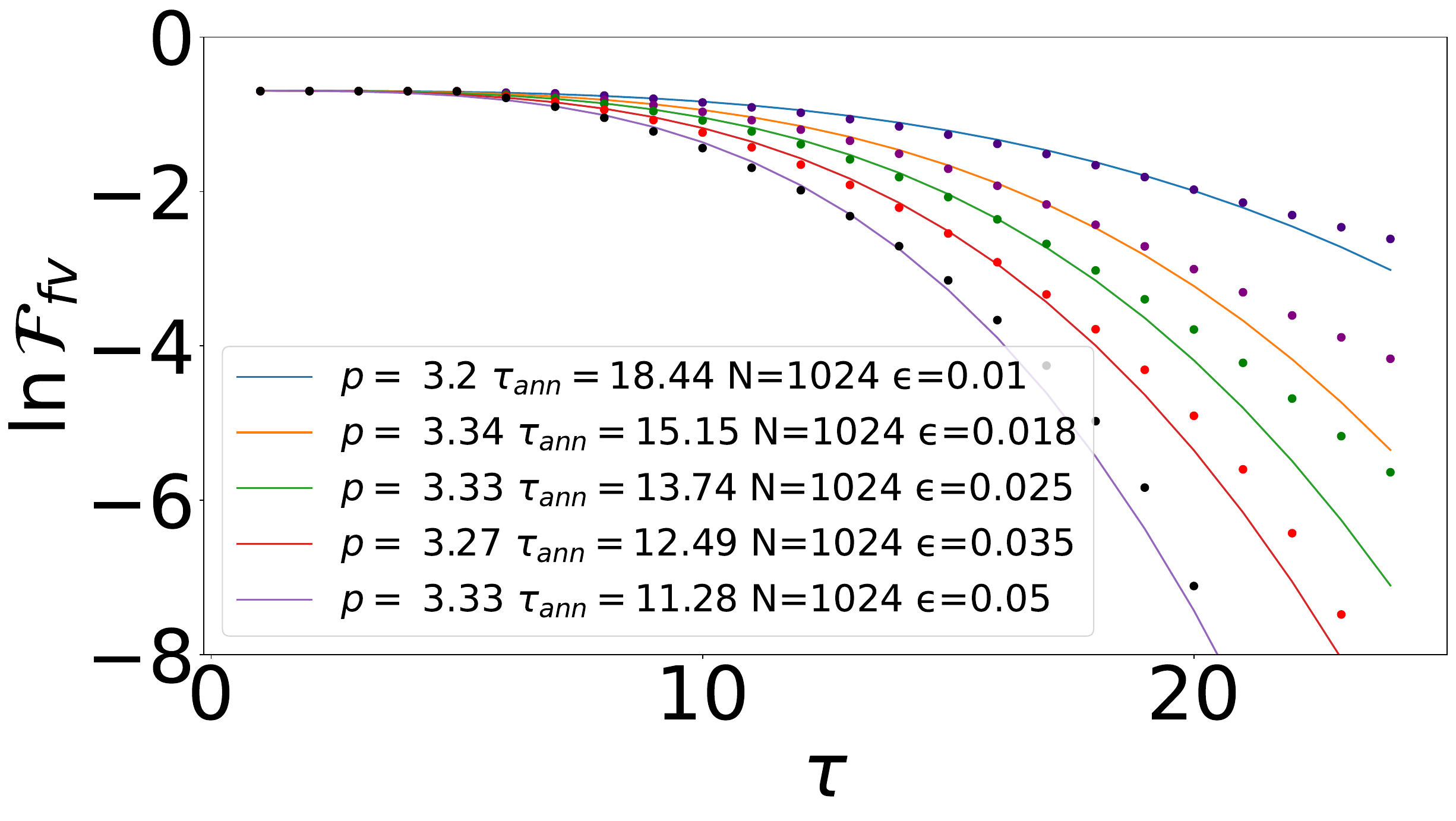}            
\end{center}
    \caption{Evolution of the false vacuum fraction ${\cal F}_{fv}$ is shown for different values of the dimensionless bias parameter $\epsilon$. The results have been obtained with $1024^3$ and $2048^3$ lattices and assuming vacuum initial conditions with the momentum cutoff $k_{cut}=1$ (top panel), $k_{cut}=5$ (bottom left panel), and $k_{cut}=0.3$ (bottom right panel). We have used Eq.~\eqref{fvfp} to fit numerical data.} \label{extended}
\end{figure}

\begin{figure}[!htb]
\begin{center}
    \includegraphics[width=0.9\textwidth]{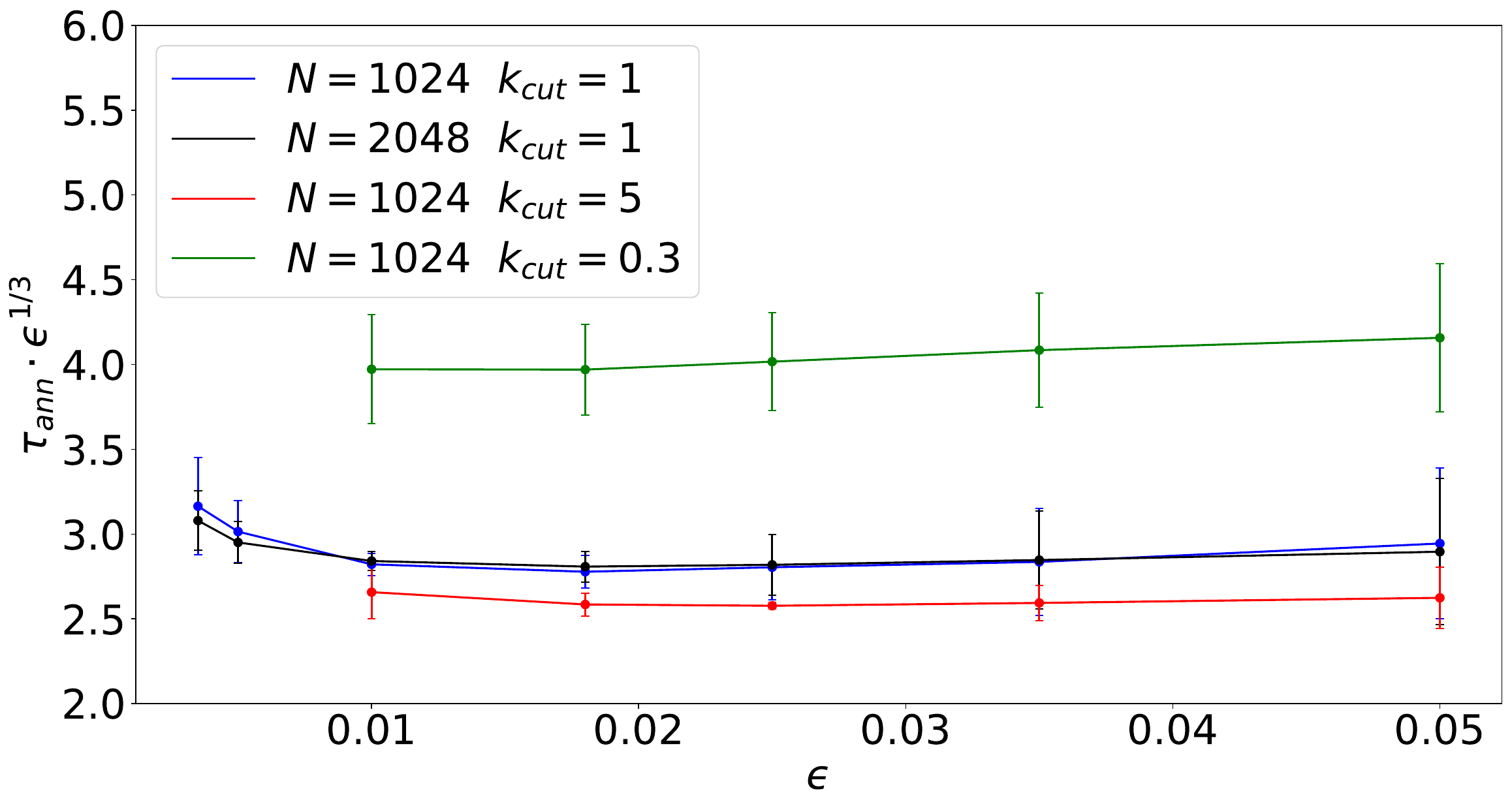}
\end{center}
    \caption{Dependence of the annihilation time $\tau_{ann}$ on the bias parameter $\epsilon$ is demonstrated for different values of the momentum cutoff $k_{cut}$ imposed on vacuum initial conditions.} \label{false_fraction}
\end{figure}

Evolution of the false vacuum fraction demonstrated in Fig.~\ref{extended} is particularly illuminating. We parameterise the false vacuum fraction as in Refs.~\cite{Kitajima:2023kzu, Pujolas}:
\begin{equation}
\label{fvfp}
{\cal F}_{fv}=\frac{1}{2} \cdot \mbox{exp} \left[-\left(\frac{\tau}{\tau_{ann}}\right)^p \right] \; ,
\end{equation}
where $p$ is assumed to be a constant. See also Refs.~\cite{Hindmarsh:1996xv, Correia:2018tty, Pujolas:2022qvs} for theoretical approaches to the problem of DW collapse. Note that the parameter $\tau_{ann}$ entering Eq.~\eqref{fvfp} matches the definition of the DW annihilation time given in Eq.~\eqref{anndef}. Compared to Ref.~\cite{Pujolas}, we observe that the power $p$ may vary as a function of the potential bias, and at this point we observe dependence on the initial conditions. Namely, the variation is rather strong for $k_{cut} =1$ and $k_{cut}=5$, while for $k_{cut}=0.3$ the parameter $p$ gets confined to a narrow range near $p = 3.3$. In this ``stabilised'' regime, values of the parameter $p$ are in a good agreement with those of Ref.~\cite{Pujolas}.

Crucially, we observe a significant departure from the commonly assumed relation $\tau_{ann} \propto 1/\sqrt{\epsilon}$. From Figs.~\ref{extended} and~\ref{false_fraction}, one infers
\begin{equation}
\label{lawdim}
\tau_{ann}\approx 3 \tau_i \cdot C (k_{cut}) \cdot  \left(\frac{\lambda v}{\epsilon} \right)^{1/3} \; ,
\end{equation}
where $C (k_{cut})$ is a coefficient, which is independent of $\epsilon$, but slightly varying with the momentum cutoff $k_{cut}$. In particular, 
one has $C (1) \approx 1$, $C(0.3) \approx 1.4$, and $C (5) \approx 0.9$. Recalling that the initial time $\tau_i$ is chosen in such a way that $H_i =1/(a_i \tau_i)=\sqrt{\lambda} v$, one can rephrase Eq.~\eqref{lawdim} in terms of the Hubble rate at radiation domination as 
\begin{equation}
\label{lawphys}
\frac{H_{ann}}{\sqrt{\lambda} v} \approx \frac{0.1}{C^2 (k_{cut})} \cdot \left(\frac{\epsilon}{\lambda v} \right)^{2/3} \; .
\end{equation}
Interestingly, the dependence $\tau_{ann} (\epsilon)$ as in Eq.~\eqref{lawdim} is consistent with the standard estimate $\rho_{wall} \sim V_{bias}$ provided that the DW energy density drops as $\rho_{wall} \propto 1/a^3$ rather than $\rho_{wall} \propto 1/a^2$. In other words, in order to satisfy the standard estimate $\rho_{wall} \sim V_{bias}$, the DW density must decrease faster than in the scenario with unbiased DWs. 
This correlates with the behavior of the scaling parameter shown in Fig.~\ref{fig:scaling} suggesting that the potential bias starts impacting DWs early in their evolution. 
Such a similarity of the behavior of biased DWs with the case of non-interacting dust particles might be not a coincidence (e.g., it may be partly due to growing amount of closed DW behaving like dust inside a Hubble patch), and it is worth investigating in the future. Notably, the observed behavior $\tau_{ann} (\epsilon)$ is largely independent of initial conditions. Namely, unlike the parameter $p$ and the coefficient $C (k_{cut})$, the relation $\tau_{ann} \propto 1/\epsilon^{1/3}$ is robust against variations of the momentum cutoff $k_{cut}$. In other words, we find that DW annihilation time is parametrically smaller (with respect to $\epsilon$) than naively expected. As we will see in the next section, this leads to drastic decrease of GW power as compared to the commonly assumed estimates.

\begin{table}[h]
    \centering
    \begin{tabular}{|c|c|c|c|c|c|c|c|}
    \hline
        $\epsilon $ &  Lattice & Reals &$k_{cut}$ & $\chi^2$, $\alpha=1/2$ & $\chi^2$, $\alpha=1/3$ & $\alpha$, best fit\\
        \hline
            $[0.01, 0.05]$  & $1024^3$ & $1$ & $0.3$ & $2.2$ & $0.05$ & $0.3$\\
            \hline
             $[0.01, 0.05]$  & $1024^3$ & $4$ & $1$ & $23.1$ & $0.29$ & $0.34$ \\
             \hline
        $[0.01, 0.05]$  & $1024^3$ & $1$ & $5$ & $4.3$ & $0.7$ & $0.34$ \\
        \hline
       $ [0.0035, 0.05]$  & $1024^3$ & $4$ & $1$ & $8.6$ & $1.1$ & $0.38$ \\ 
       \hline
         $[0.0035, 0.05]$  & $2048^3$ &  $1$ & $1$ & $42.2$ & $1.9$& $0.37$ \\
         \hline
    $[0.0009, 0.0016]$  & $3240^3$  &   $4$ & $-$ & $5.0$ & $0.05$ & $0.32$ \\ 
    \hline
        $[0.0008, 0.0016]$  & $3240^3$  &   $4$ & $-$ & $3.0$ & $1.1$ & $0.42$ \\ 
    \hline
    \end{tabular}
       \caption{The power-law behaviour $\tau_{ann} \propto 1/\epsilon^{\alpha}$ has been studied using the $\chi^2$ analysis for the cases $\alpha=1/2$ and $\alpha=1/3$. The results are demonstrated for a selection of lattices at different resolutions/for different initial conditions of the field $\chi$/for different number of realizations of the initial scalar field configuration. The best fit value $\alpha$ is also shown in the Table. Numbers in the last two lines 
       have been filled using the results of simulation data from Ref.~\cite{Pujolas}. The cutoff scale on vacuum initial conditions is not specified there.}\label{chi2}
\end{table}

To derive the relation~\eqref{lawdim} we have probed the ranges $\epsilon =[0.0035, 0.05]$ and $\epsilon=[0.01, 0.05]$ using $2048^3$ and $1024^3$ lattices, respectively. We have performed $\chi^2$-analysis to test the hypothesis $\tau_{ann} \propto 1/\epsilon^{1/3}$ 
versus $\tau_{ann} \propto 1/\epsilon^{1/2}$. Results shown in Table~\ref{chi2} demonstrate a clear preference for the former 
hypothesis over the latter one. Steepening of the behavior $\tau_{ann} (\epsilon)$ in the range of very small $\epsilon$ is likely attributed to 
the fact that in this case the network decay takes place at sufficiently late times, when effects, related to finite lattice size/resolution, become significant. On the other hand, in this range the outcome is less dependent on the possible impact of initial conditions, 
and therefore the steepening may indicate instead that the commonly assumed law $\tau_{ann} \propto 1/\epsilon^{1/2}$ is restored at very small $\epsilon$.

To understand 
this issue better, we have analysed the results of numerical simulations for the false vacuum decay from Ref.~\cite{Pujolas}, where the range $\epsilon =[0.0008, 0.0016]$ has been explored with higher resolution simulations. The corresponding $\chi^2$-values are also demonstrated 
in Table~\ref{chi2}. We again observe preference of the behaviour $\tau_{ann} \propto 1/\epsilon^{1/3}$ over $\tau_{ann} \propto 1/\epsilon^{1/2}$. Simulations of Ref.~\cite{Pujolas} also exhibit a significant steepening of the behaviour $\tau_{ann} (\epsilon)$, 
if one allows for large simulation times, as in our case. As the ranges of $\epsilon$ 
considered in this work and in Ref.~\cite{Pujolas} do not overlap, this feature is perhaps a numerical artefact. Nevertheless, it serves as a motivation for further studies 
of the behaviour $\tau_{ann} (\epsilon)$ using high resolution simulations and for a larger scope of initial conditions.

\begin{figure}[!htb]
\begin{center}
    \includegraphics[width=0.49\textwidth]{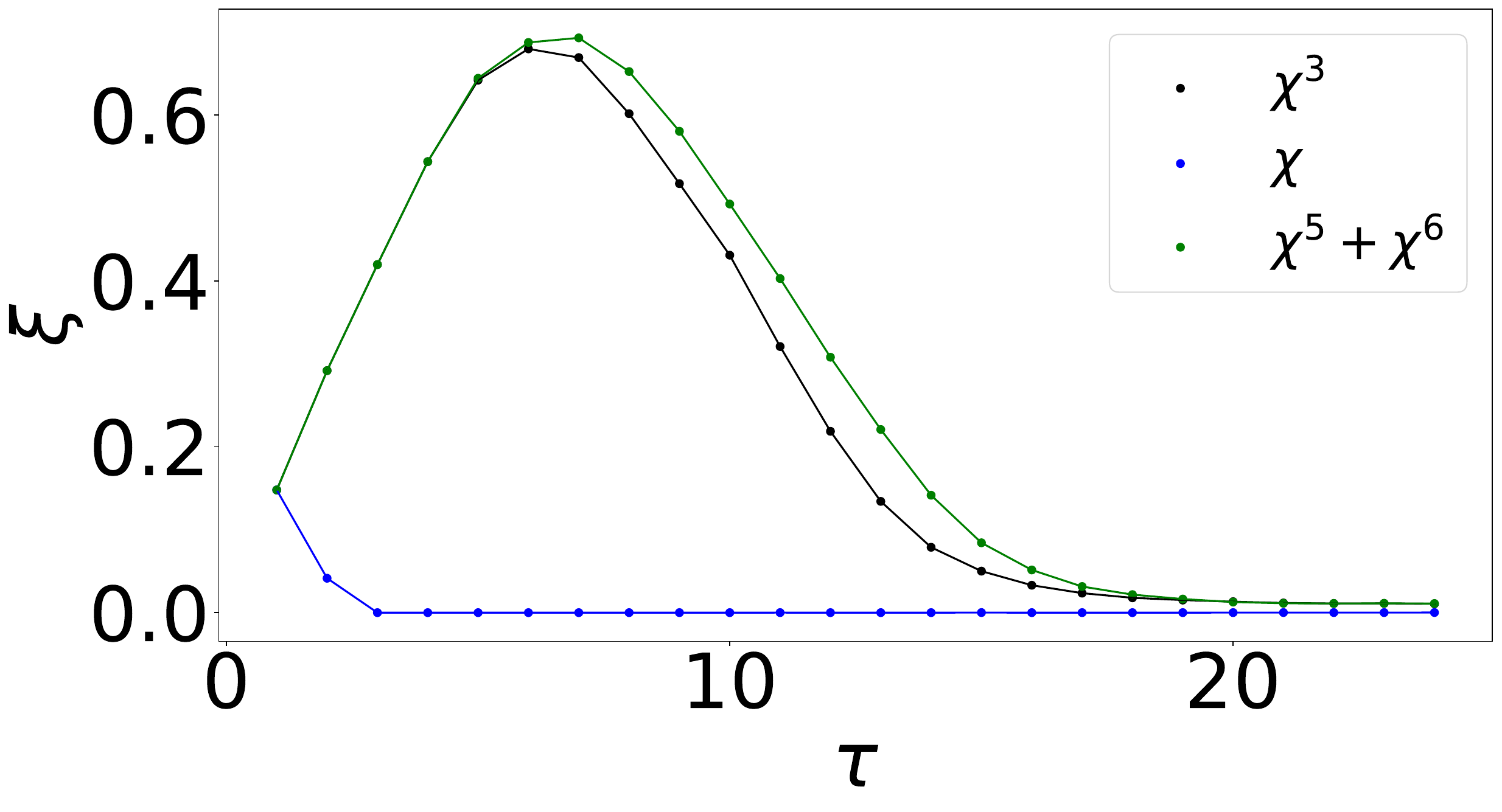} 
    \includegraphics[width=0.49\textwidth]{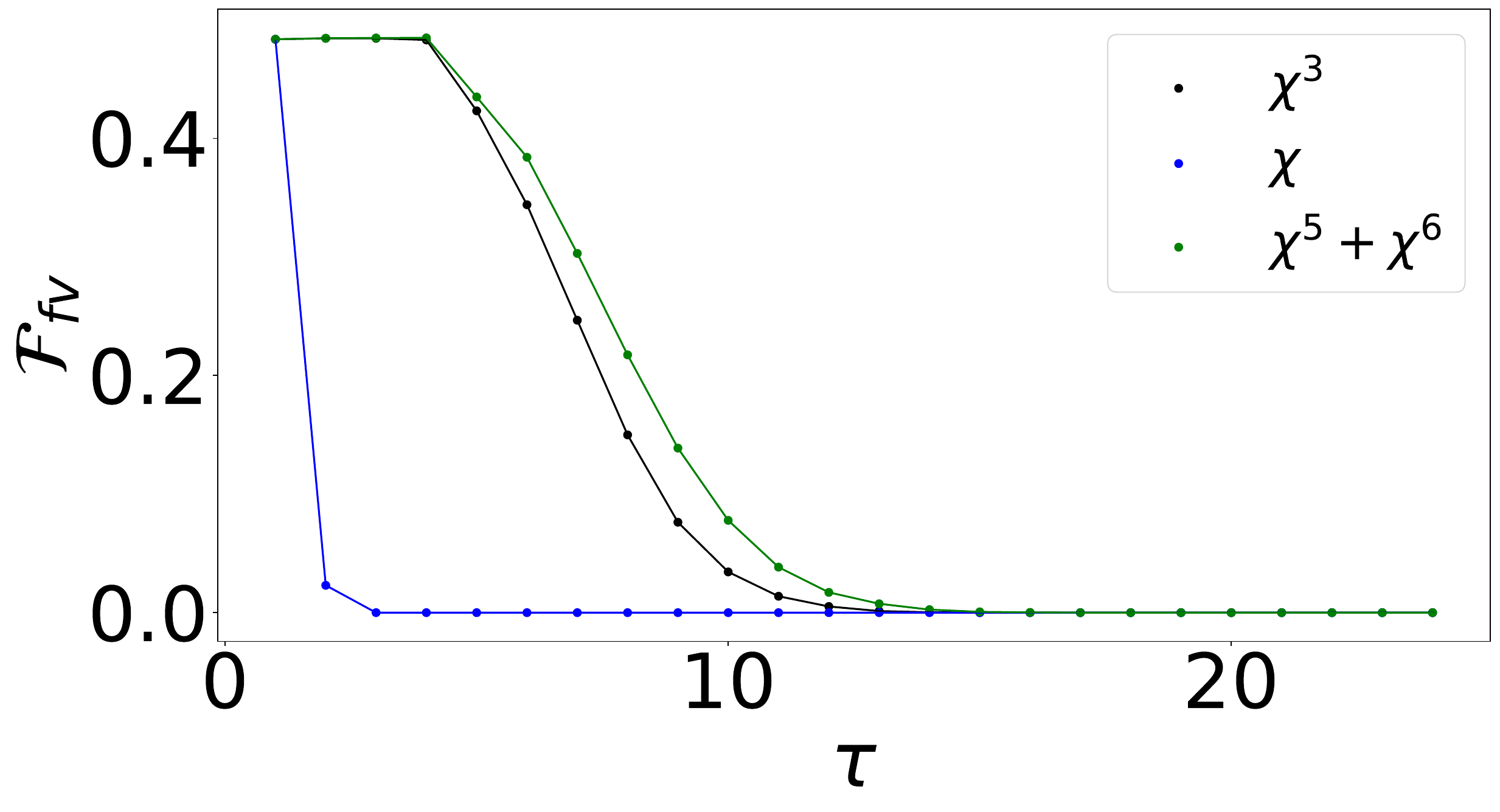} 
\end{center}
    \caption{Comparison of the potential bias impact on DW formation and evolution caused by the symmetry breaking potentials $V_{breaking}=\epsilon \chi^3$, $\epsilon \chi v^2$, and $\epsilon (\chi^5/v^2+\chi^6/v^3)$. The bias parameter $\epsilon$ has been set to $\epsilon=0.05$ (in units of $\lambda v$). The impact is demonstrated in terms of the scaling parameter $\xi$ (left panel) and the false vacuum fraction ${\cal F}_{fv}$ (right panel). Simulations have been carried out with $1024^3$ lattice. DWs are not formed in the case of a linear potential bias (for the choice of $\epsilon$ assumed).} \label{3_5}
\end{figure}

Let us consider other choices of $Z_2$-symmetry breaking caused by the linear term $V_{breaking}=\epsilon \chi v^2$ as well as the non-renormalisable term of the form $V_{breaking}=\epsilon \chi^5/v^2+\epsilon' \chi^6/v^3$. In the latter case, the term $\epsilon' \chi^6/v^3$ with some positive constant $\epsilon'$ has been added to keep the overall potential bounded from below. As it is clear from Fig.~\ref{3_5}, the impact of the higher order potential bias on DW evolution is almost indistinguishable from that of the cubic bias. On the other hand, the effect of the linear bias is dramatically different. The likely reason is that the linear bias term changes behavior of its potential $V(\chi)$ around the minimum $\chi=0$ leading to non-zero $\partial V/\partial \chi$ at $\chi=0$. As a result, one of the minima becomes slightly preferred meaning that 
the linear potential bias triggers the population bias. 
This may explain, why the DW network is not formed in this case, cf. Ref.~\cite{Krajewski:2021jje}. In this work we focus on the effects, which are entirely due to the potential bias, and omit any further discussion of the case $V_{breaking} \propto \chi$.

\section{Numerical results: gravitational waves}
\label{sec:gw}

\begin{figure}[!htb]
\begin{center}
\includegraphics[width=\textwidth]{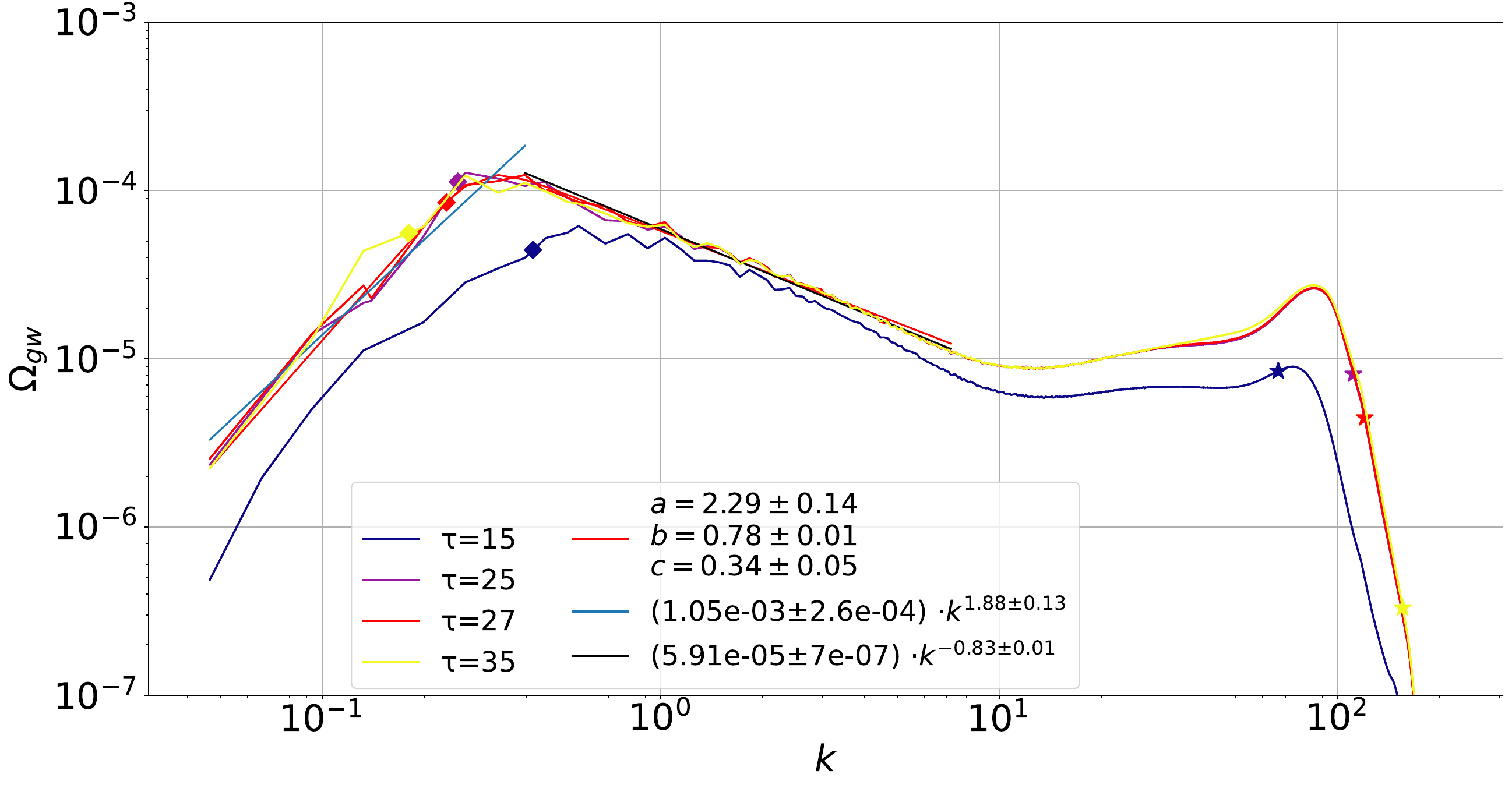} 
\includegraphics[width=\textwidth]{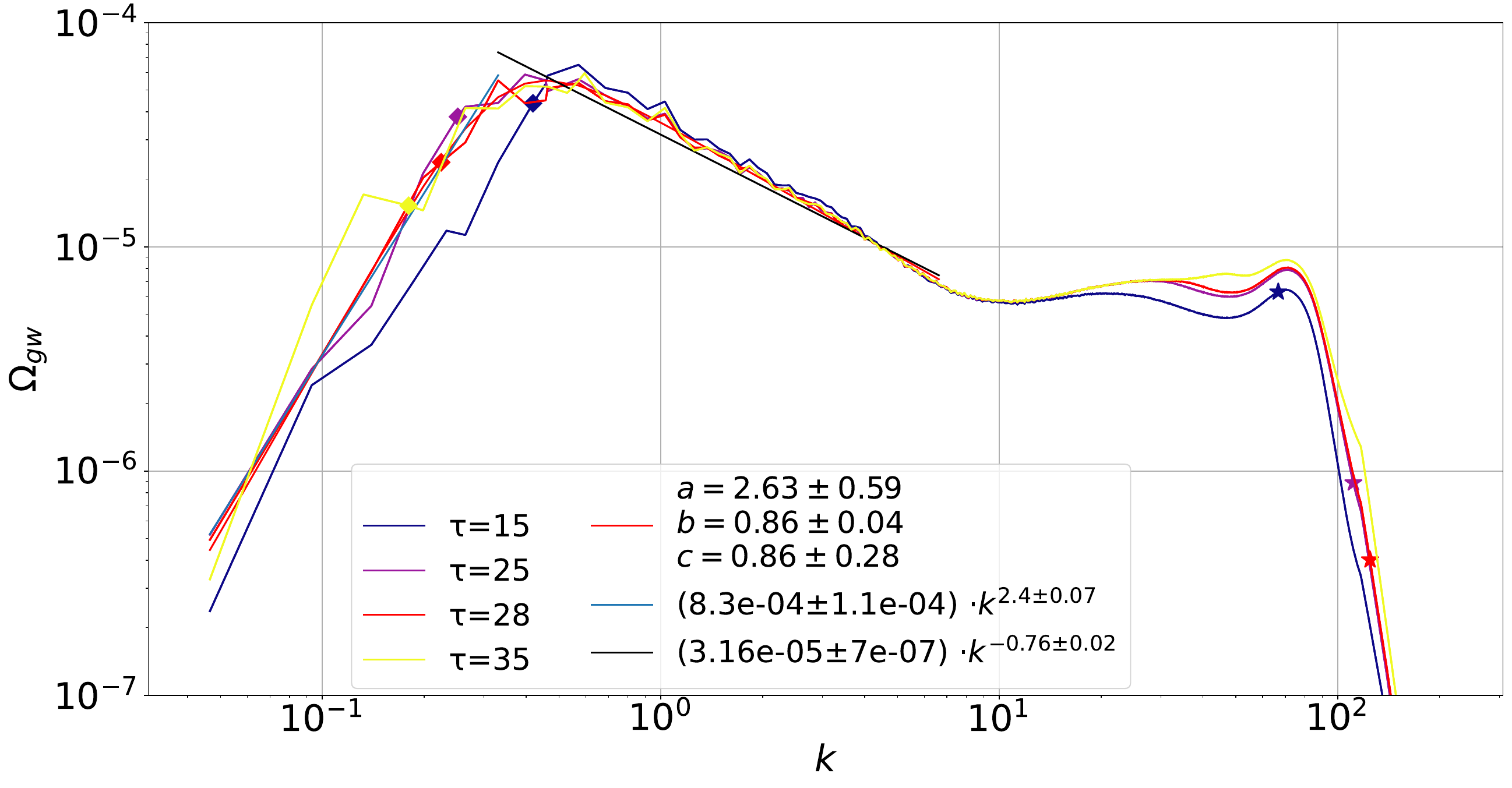} 
\end{center}
    \caption{GW spectra obtained with the $2048^3$ lattice for the bias parameter $\epsilon=0.025$ (top panel) and $\epsilon=0.05$ (bottom panel) starting with vacuum initial conditions and momentum cutoff $k_{cut}=1$. Shown also are power-law fits and the fit described by Eq.~\eqref{fitting}.} \label{spectra_combined_0025}
\end{figure}

\begin{figure}[!htb]
\begin{center}
\includegraphics[width=\textwidth]{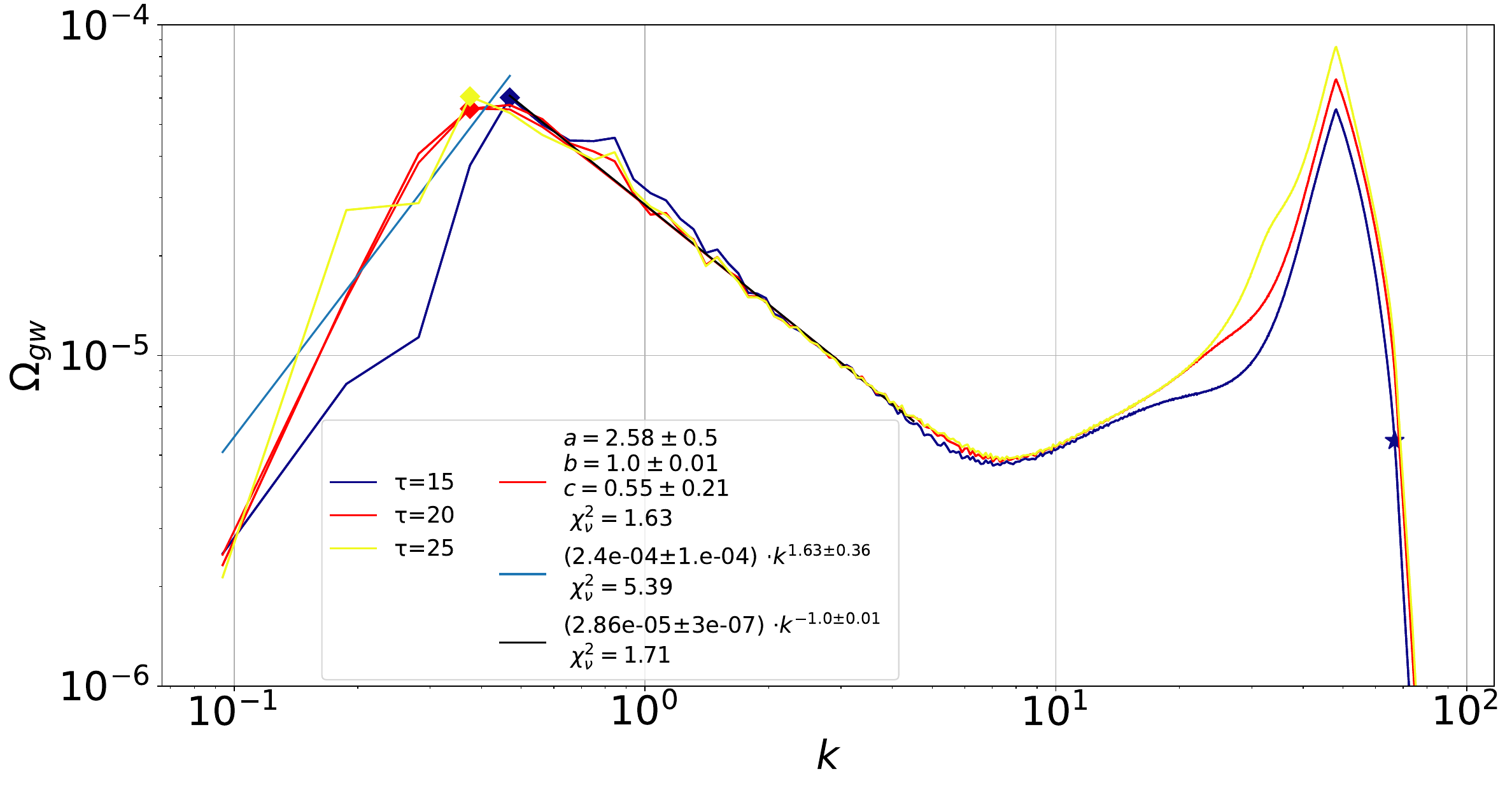} 
\end{center}
    \caption{GW spectra obtained by averaging over 10 realisations assuming the bias parameter $\epsilon=0.05$ and vacuum initial conditions with momentum cutoff $k_{cut}=1$. Simulations 
    have been performed with $1024^3$ lattice. The spectra have been produced with the fixed ratio $\kappa'/\kappa=\pi/2$ (see description in the end of Sec.~\ref{sec:nsetup}). This explains appearance of the artificial UV peak compared to Fig.~\ref{spectra_combined_0025}. Shown also are power-law fits and the fit described by Eq.~\eqref{fitting}. The quality of the fit to the IR slope is not good due to a lack of points.} \label{spectra_10_reals}
\end{figure}

\begin{figure}[!htb]
\begin{center}
     \includegraphics[width=0.45\textwidth]{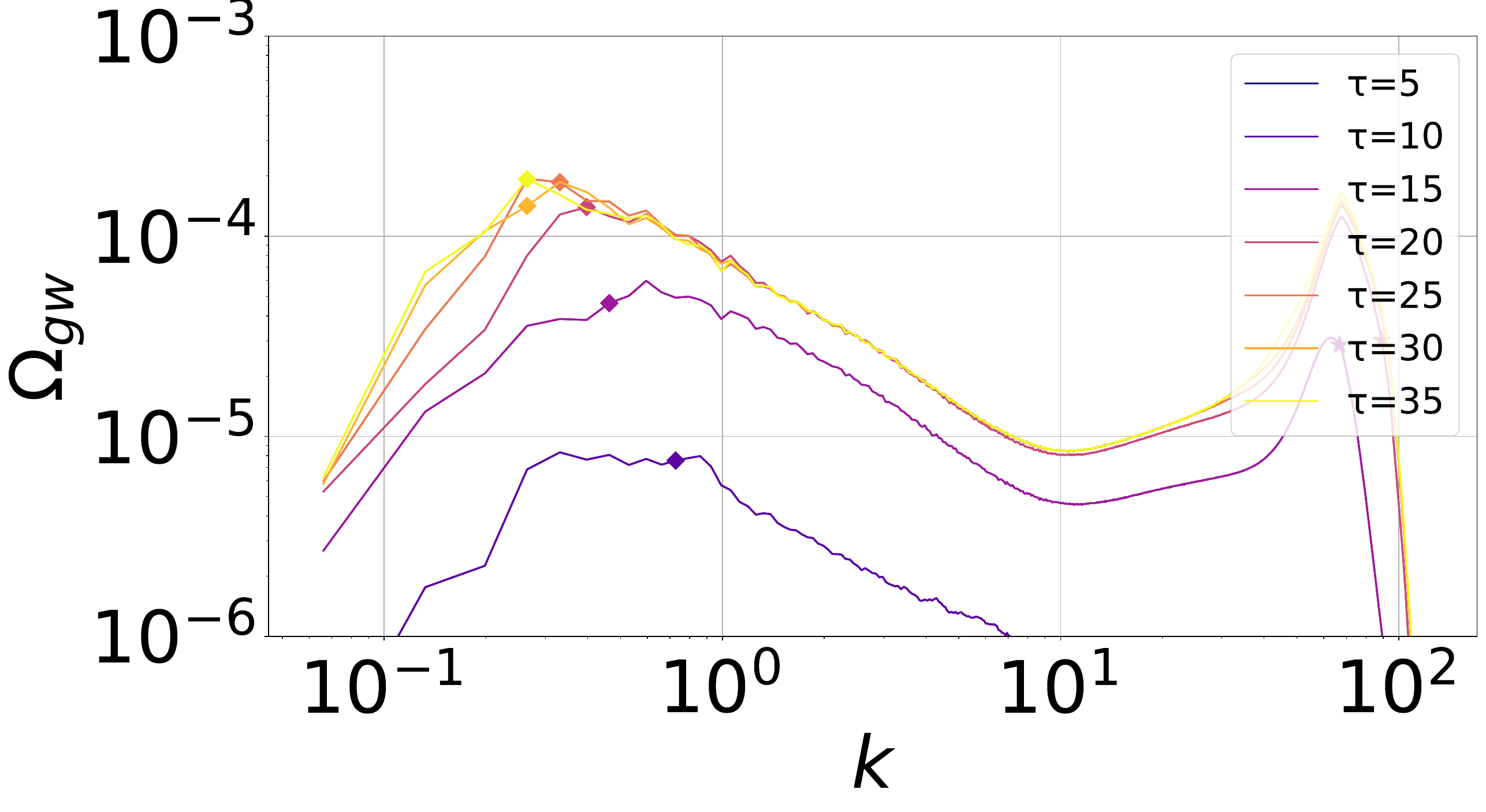} 
    \includegraphics[width=0.45\textwidth]{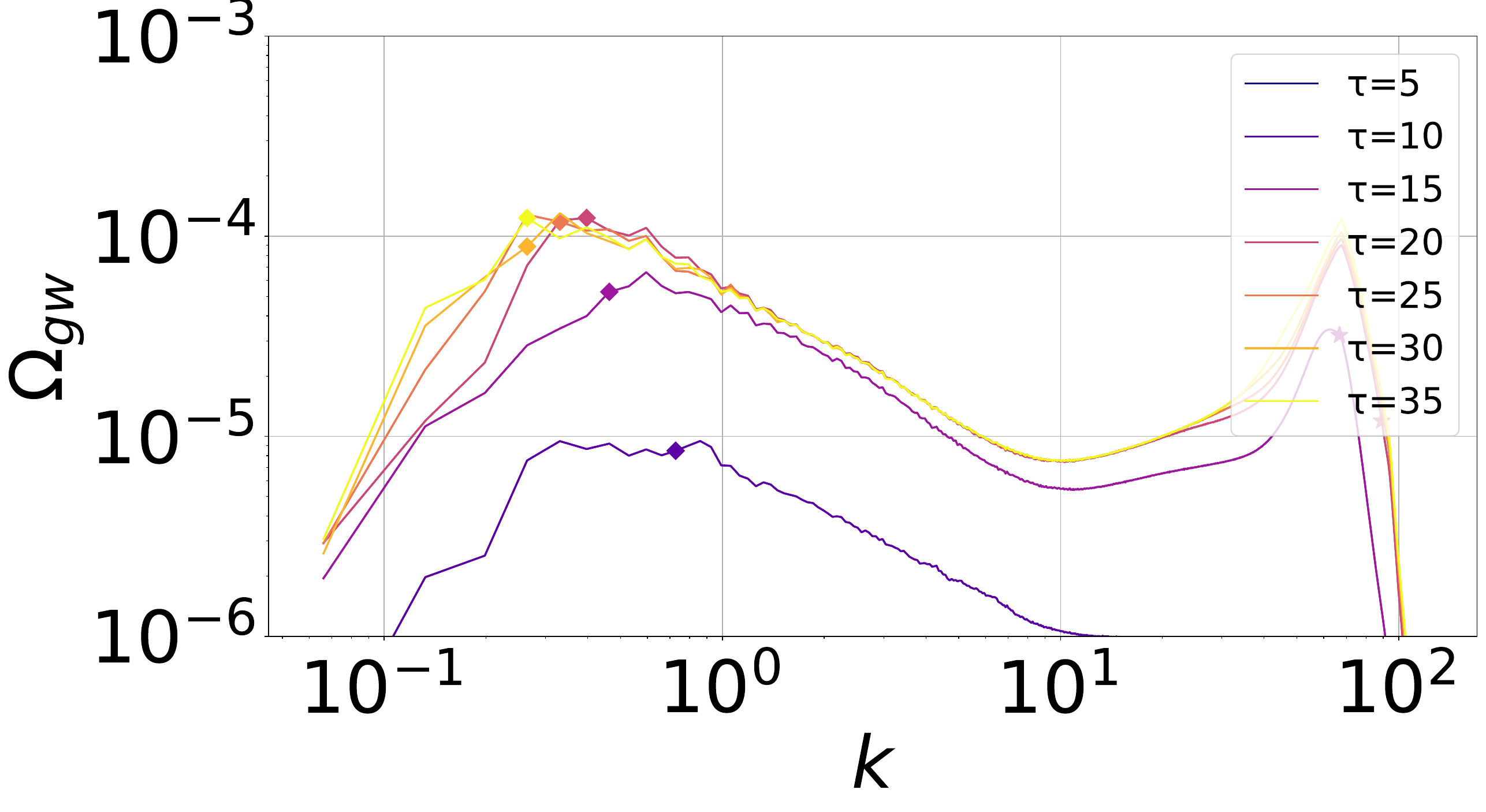} 
        \includegraphics[width=0.45\textwidth]{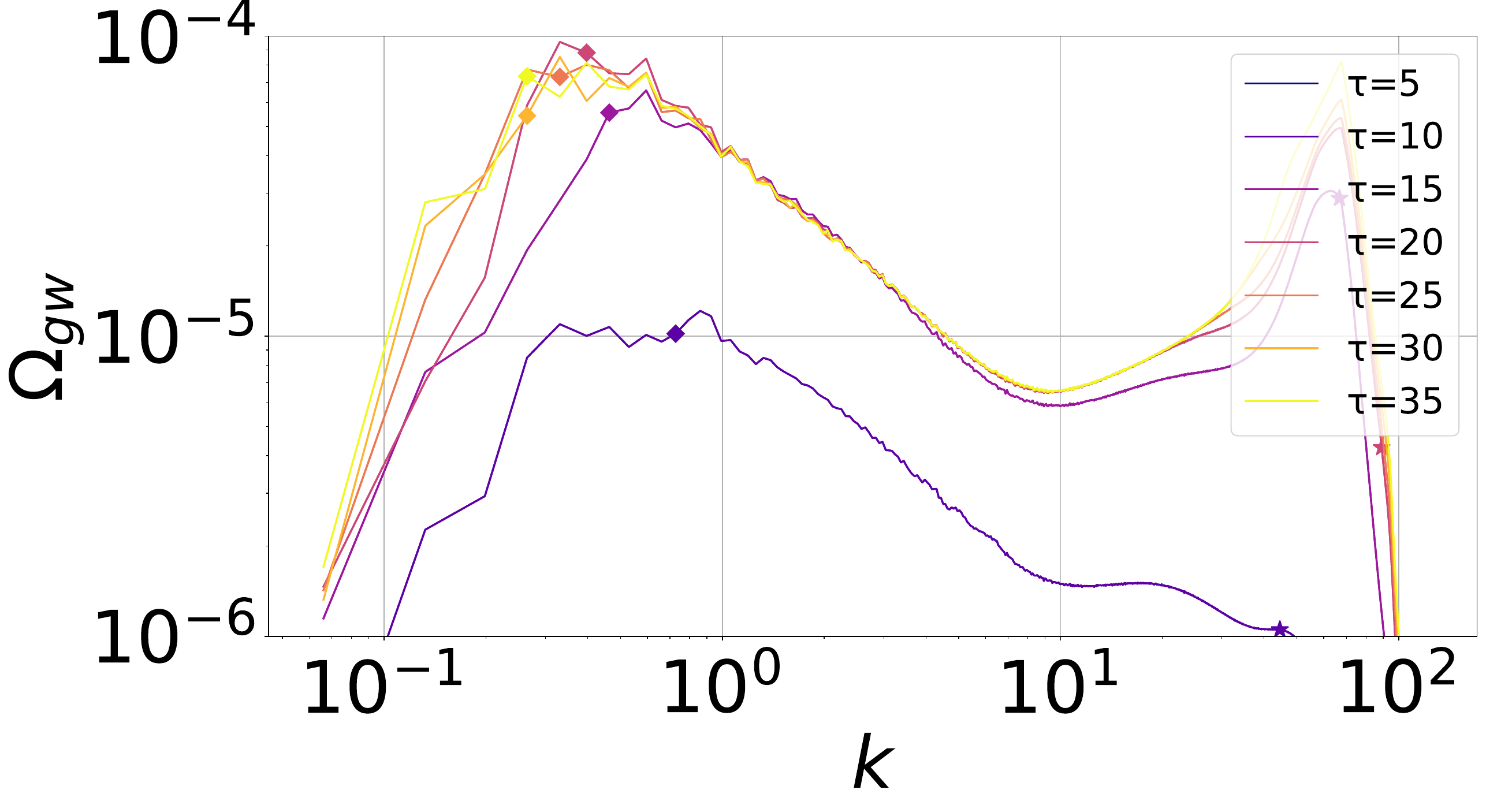} 
    \includegraphics[width=0.45\textwidth]{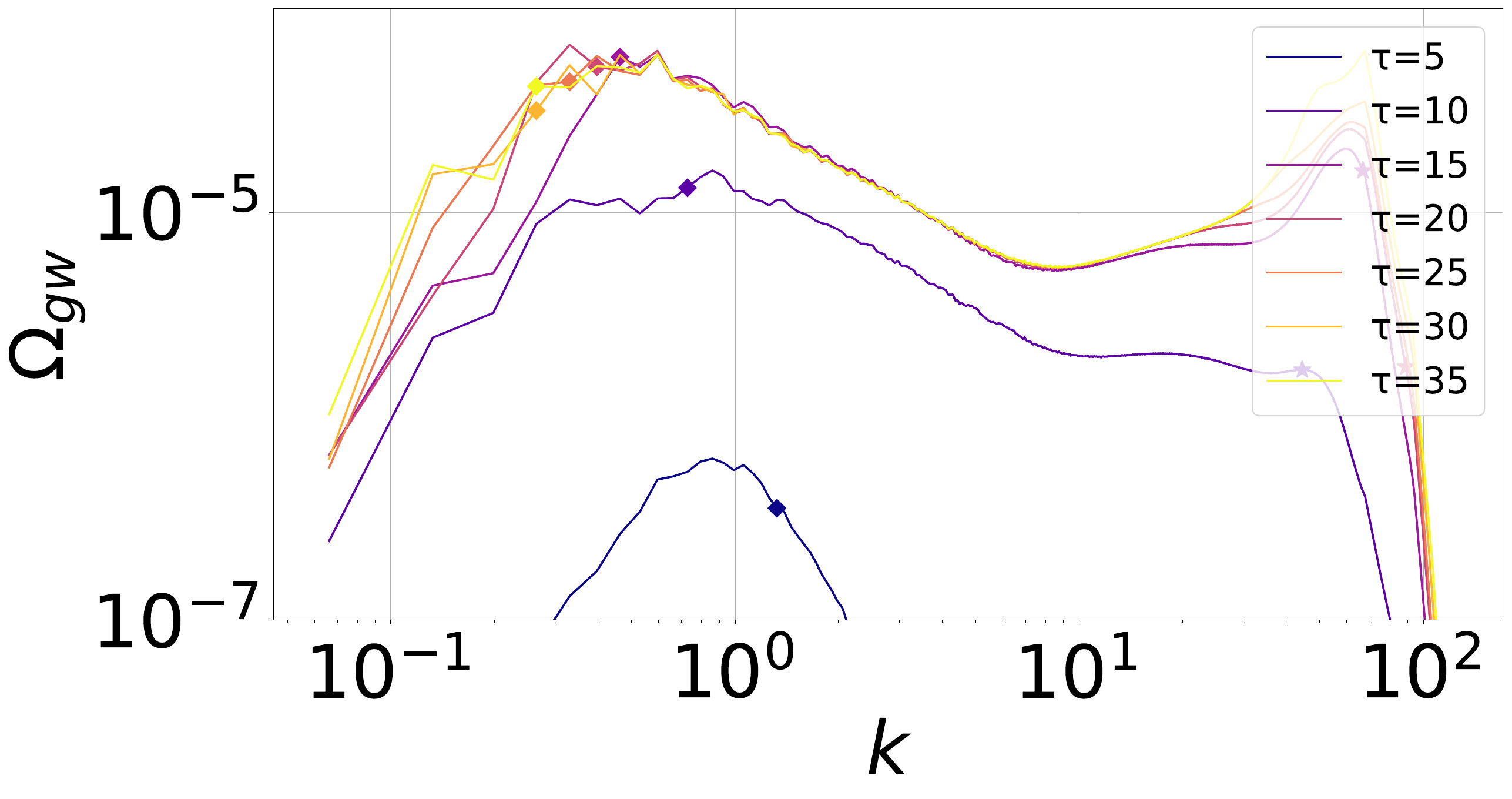} 
\end{center}
    \caption{GW spectra obtained at $2048^3$ lattice with vacuum initial conditions and momentum cutoff $k_{cut}=1$. The bias parameter equals  
    $\epsilon=0.018$ (top left panel), $\epsilon=0.025$ (top right panel), $\epsilon=0.035$ (bottom left panel), $\epsilon=0.05$ (bottom right panel). The spectra are produced with the fixed ratio $\kappa'/\kappa=\pi/2$ (see description in the end of Sec.~\ref{sec:nsetup}). This explains appearance of the artificial UV peak compared to Fig.~\ref{spectra_combined_0025}.} \label{spectra_various}
\end{figure}

Results of numerical simulations of GWs production by biased DWs with vacuum initial conditions are shown in Figs.~\ref{spectra_combined_0025},~\ref{spectra_10_reals}, and~\ref{spectra_various}. We restrict our analysis to relatively large values $\epsilon=0.018, 0.025, 0.035, 0.05$, because otherwise it takes a long time for the DW network to collapse and hence for the GW spectrum to get stabilised. We have fixed other model parameters to be 
$\lambda=0.03$, $v=6 \cdot 10^{16}~\mbox{GeV}$, and the number of relativistic degrees of freedom in the primordial plasma $g_* (T)=100$ 
(we keep it constant within the relevant time span of GW production). The results can be rescaled to arbitrary values, as we demonstrate below. Figures~\ref{spectra_combined_0025} and~\ref{spectra_various} show spectra obtained with one simulation each on $2048^3$ lattice, while the plot on Fig.~\ref{spectra_10_reals} is produced by averaging over 10 simulations performed with different seeds for initial values of $\chi$ on $1024^3$ lattice. 
\begin{figure}[!htb]
\begin{center}
     \includegraphics[width=0.9\textwidth]{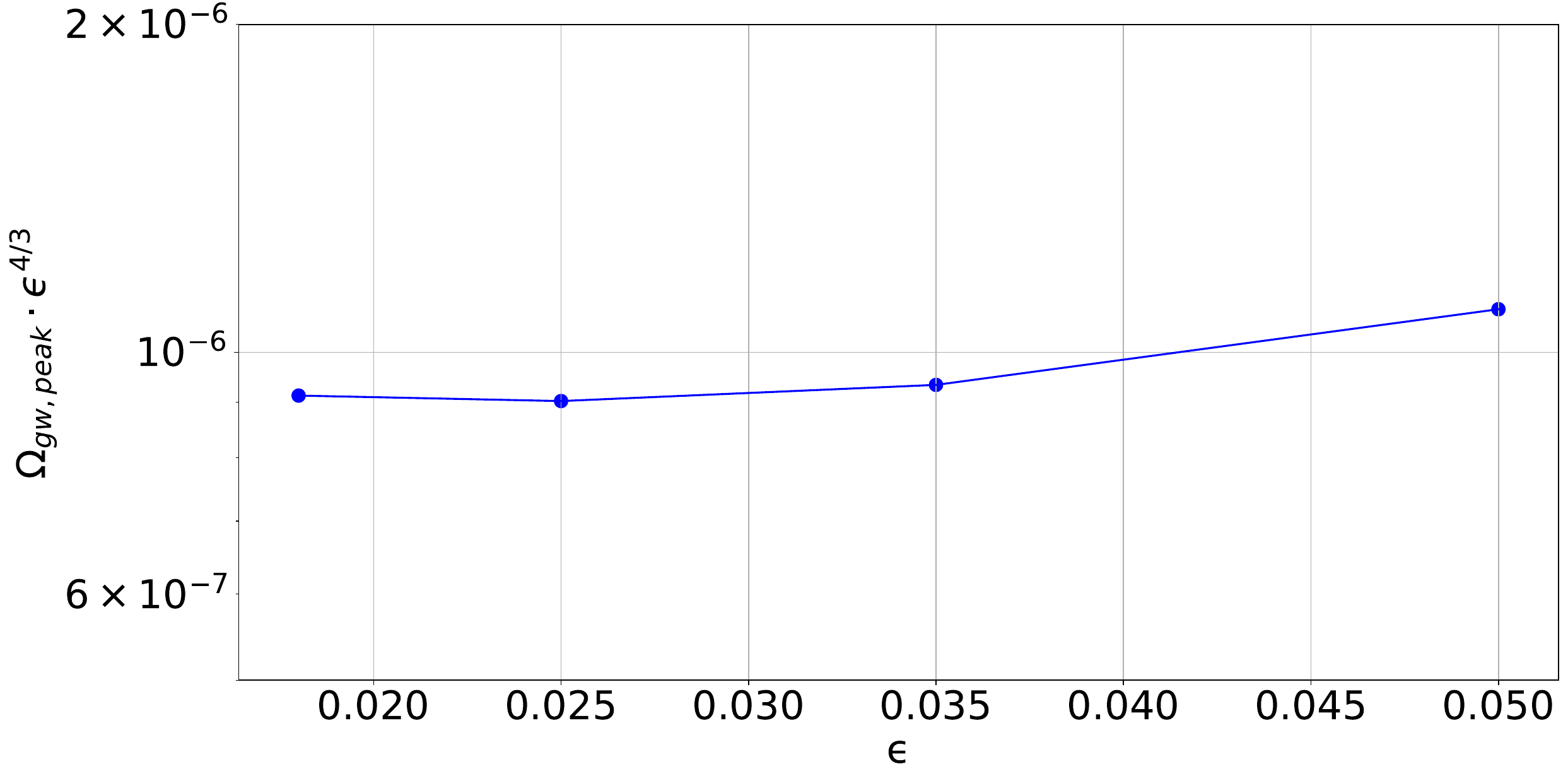} 
\end{center}
    \caption{Demonstrated is dependence of $\Omega_{gw, peak}$ on the bias parameter $\epsilon$ extracted from Fig.~\ref{spectra_various}. Vacuum initial conditions with momentum cutoff $k_{cut}=1$ have been assumed.} \label{omega_peak}
\end{figure}

In this way, one can conservatively estimate error bars related to the statistical nature of the initial field $\chi$ distribution. 
The spectrum of GWs in Fig.~\ref{spectra_combined_0025} is obtained with the procedure described in Sec.~\ref{sec:nsetup}, 
cf. Ref.~\cite{Dankovsky:2024zvs}. Namely, one varies the ratio $\kappa'/\kappa$ according to Eq.~\eqref{zoom}, with parameters $\kappa'$ and $\kappa$ controlling IR and UV physics, respectively. This allows us to ``zoom in'' the IR, near peak, and UV parts of the spectrum. Then one combines different parts to get the full spectrum. The procedure is designed to get rid of/mitigate the effect of non-physical factors. In particular, as it is clear from Figs.~\ref{spectra_10_reals} and~\ref{spectra_various} obtained with fixed $\kappa'/\kappa=\pi/2$, one observes the strong UV peak. The latter has a non-physical origin, as it has been proven in Ref.~\cite{Dankovsky:2024zvs}: its location changes depending on the lattice resolution.

Let us first discuss characteristics of GWs in the peak region. The peak frequency of GWs is given by 
\begin{equation}
\label{kpeak}
\frac{k_{peak}}{2\pi a_{ann}} \simeq 0.6 H_{ann} \; ,  
\end{equation}
which matches well the result obtained in the case of unbiased DWs~\cite{Dankovsky:2024zvs}. The fractional energy density of GWs at maximum reconstructed from Figs.~\ref{spectra_combined_0025} and~\ref{spectra_various} is given by   
\begin{equation}
\label{gwpeak}
\Omega_{gw, peak} \simeq 1.3 \cdot 10^{-8} \left(\frac{H_{i}}{H_{ann}} \right)^2  \cdot \left(\frac{v}{6 \cdot 10^{16}~\mbox{GeV}}\right)^4 \; .
\end{equation}
Note that the dependence on the expectation value $v$ and the constant $\lambda$, encoded in $H_i =\sqrt{\lambda} v$, follows from the relation $\rho_{gw} \propto \sigma^2_{wall}/M^2_{P}$. The factor $\sim H^2_{ann}$ in the denominator comes from the Friedmann equation, $\rho_{tot}=3H^2_{ann} M^2_{P}$, which relates the total energy density entering $\Omega_{gw}$ and the Hubble parameter. 
The peak energy density of GWs exceeds by 
about an order of magnitude the naive expectations, where $\tau_{ann}$ would be taken as the last instant of GW emission, cf. Eq. (58) of Ref.~\cite{Dankovsky:2024zvs}.
The deviation can be explained by the fact that GW production continues some time after the DW collapse $\tau_{ann}$. Namely, one has 
\begin{equation}
\tau_{gw} \sim 2 \tau_{ann} \; ,
\end{equation}
where $\tau_{gw}$ is the time of peak production of GW energy, cf. Refs.~\cite{Kitajima, Pujolas}. Indeed, one can see in Figs.~\ref{spectra_combined_0025} and~\ref{spectra_various}, that the growth of GW energy density relative to radiation 
continues beyond the time $\tau_{ann}$. The possible physical reason is that the inhomogeneous particle distribution following the DW collapse continues to source GW waves. 
Such a source is terminated, once the spatial distribution of particles becomes homogeneous. If particles are relativistic, 
it takes about one Hubble time for the particle distribution to become homogeneous; since this time on the source of GWs is terminated, and their growth stops.

\begin{figure}[!htb]
\begin{center}
    \includegraphics[width=\textwidth]{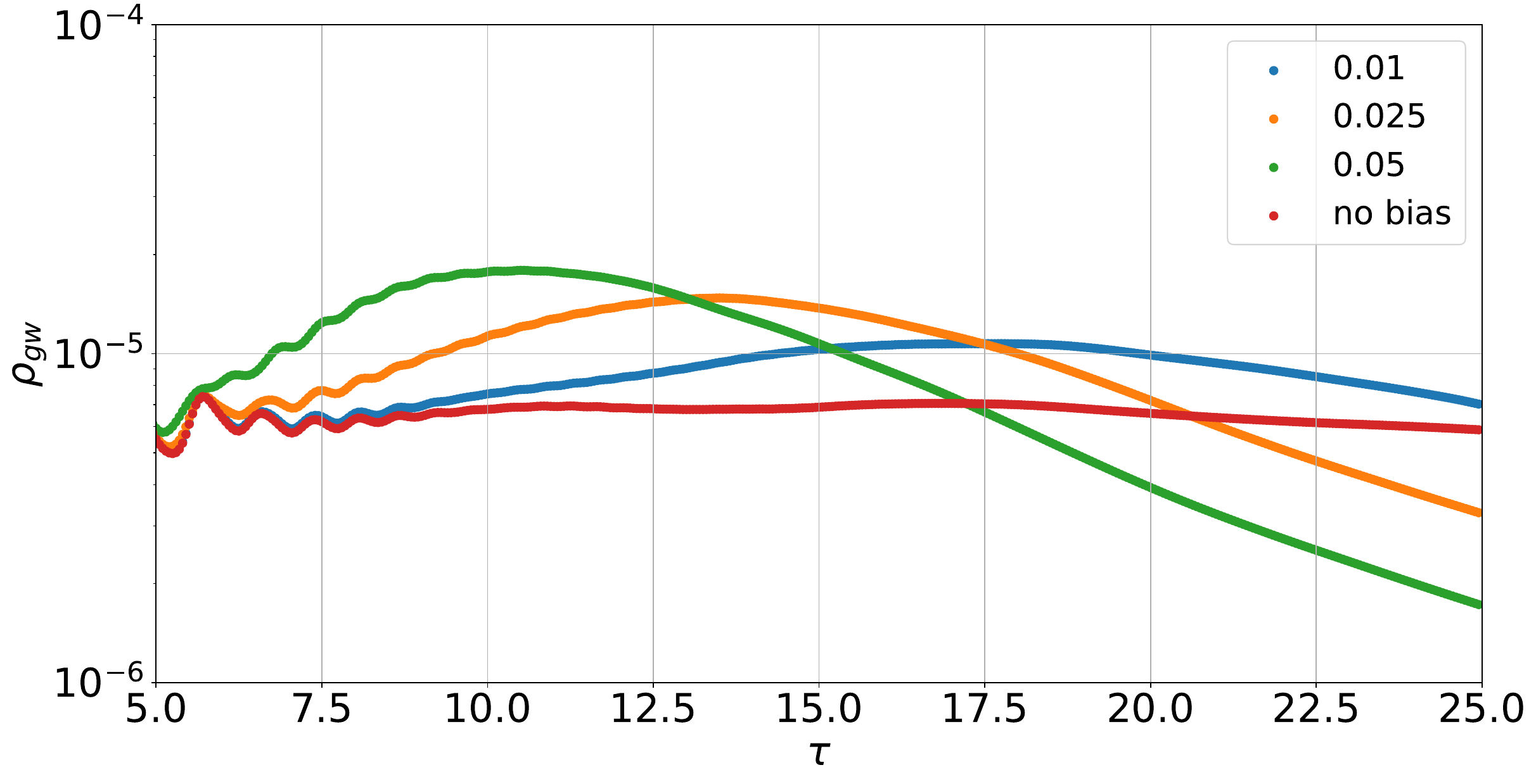} 
\end{center}
    \caption{The energy density of GWs obtained for a selection of bias parameters $\epsilon$ assuming vacuum initial conditions with momentum cutoff $k_{cut}=1$. Simulations have been carried out with $1024^3$ lattice.} \label{gw_total}
\end{figure}

One important comment is in order here. To write Eq.~\eqref{gwpeak}, we have used $\rho_{gw} \propto \sigma^2_{wall}/M^2_{P}$, which assumes that DWs are in the scaling regime. Note, however, that the scaling law is generically violated once the potential bias is included, and this violation is particularly prominent close to the moment of DW annihilation. Such a violation may entail an additional dependence on the 
bias parameter $\epsilon$. This dependence is, however, expected to be rather soft, as one can see from Fig.~\ref{gw_total}.

Substituting Eq.~\eqref{lawphys} into Eqs.~\eqref{kpeak} and~\eqref{gwpeak}, we can rephrase the peak frequency and the fractional energy density of GWs at maximum in terms of model parameters. We also take into account the redshift and write the present day values of the corresponding quantities: 
\begin{equation}
\label{fpeak}
f_{peak} \simeq 8~\mbox{nHz} \, \lambda^{1/4} \cdot \left( \frac{\epsilon}{10^{-36}\cdot \lambda v} \right)^{1/3}   \cdot \sqrt{\frac{v}{100~\mbox{TeV}}} \cdot \left(\frac{100}{g_* (T_{ann})} \right)^{1/12} \; ,
\end{equation}
and
\begin{equation}
\label{gwpeak1}
\Omega_{gw, peak} h^2_0 \simeq 1 \cdot 10^{-10} \cdot \left(\frac{v}{100~\mbox{TeV}} \right)^4 \cdot \left( \frac{10^{-36} \cdot \lambda v}{\epsilon} \right)^{4/3} \cdot \left(\frac{100}{g_* (T_{ann})} \right)^{1/3} \; . 
\end{equation}
Note that we have set $C (k_{cut}) \approx 1$ in Eq.~\eqref{lawphys}, because we assume the momentum cutoff $k_{cut}=1$ for vacuum initial conditions. The reference values in Eqs.~\eqref{fpeak} and~\eqref{gwpeak1} are chosen for convenience of comparison with the 
recent PTA findings~\cite{NANOGrav:2023gor, NANOGrav:2023hvm, EPTA:2023fyk, EPTA:2023xxk, Xu:2023wog, Reardon:2023gzh}.   
As it follows from Eq.~\eqref{gwpeak}, see also Fig.~\ref{omega_peak}, dependence of $\Omega_{gw, peak}$ on $\epsilon$ is considerably softer compared to the commonly assumed 
$\Omega_{gw, peak} \propto 1/\epsilon^2$. This is due to the earlier annihilation time revealed in our calculations as compared to the naively expected annihilation time. Consequently, one gets a much weaker GW signal for the same $\epsilon$, given that the latter is typically taken to be extremely small, as it is indicated in Eq.~\eqref{gwpeak1}. This conclusion of our analysis is at odds with those of Refs.~\cite{Cyr:2025nzf, Pujolas, Notari:2025kqq}.

Now let us discuss the spectral shape of GWs. Spectral indices in Figs.~\ref{spectra_combined_0025} and~\ref{spectra_10_reals} have been obtained using the following procedure. Assuming the power-law fit, so that $\ln \Omega_{gw} = \ln \alpha + \beta \ln k$ with constant $\alpha$ and $\beta$, we obtain values of $\alpha$ and $\beta$ and their errors using the linear least squares method. To estimate the spectral index in the IR part of the spectrum we take all the points from $k_{min}=2\pi/L$ to $k_{peak}$. In the UV part of the spectrum, we take all the points from $k_{peak}$ to the momentum $k$, which is $1.5$ times smaller than the IR edge of the plateau. All the fits have been performed using function \texttt{curve\_fit} function from Python library \texttt{SciPy}. The choice of conformal times to fit GW spectra is dictated by the following considerations: the spectra should be stabilised at those times, and oscillatory features imposed on the spectra should not be prominent. Note that oscillations must be averaged out anyway, when calculating the GW energy density. Such an averaging can be trivially implemented, if the GW period is small relative to the Hubble radius. Due to a limited time span of our simulations, however, certain modes of interest have wavelengths of the order of the horizon radius by the end of simulations, and it may be difficult to get rid of oscillations in the IR part of the spectrum. In this situation, one mitigates oscillations simply by picking the spectrum at the time, when oscillatory effects are accidentally small.

The GW spectrum can be split into four regions: 

i) The IR part of the spectrum, where one has $\Omega_{gw} \propto f^3$ as $f \rightarrow 0$ for causality considerations. Departures from this asymptotic behavior in Figs.~\ref{spectra_combined_0025},~\ref{spectra_10_reals}, and~\ref{spectra_various} are due to the limited range of momenta available in simulations, i.e., $k \geq 2\pi/L$.

ii) The close-to-maximum UV part has the spectral shape approximately described by $\Omega_{gw} \propto f^{-1}$, which agrees with the results obtained recently in Ref.~\cite{Cyr:2025nzf}, but disagrees with those of Ref.~\cite{Notari:2025kqq}.
Here one observes a rather strong deviation from the unbiased case characterised by the spectral shape $\Omega_{gw} \propto f^{-1.5}$~\cite{Dankovsky:2024zvs}. We attribute this feature to the richer substructure in the form of closed DWs in the case of biased DWs. Indeed, it has been observed in Refs.~\cite{Dankovsky:2024zvs, Garagounis:2002kt} that the unbiased DW network is dominated by a single ``infinite'' wall, while the number of small closed walls is negligible. It is interesting to note that the spectral shape $\Omega_{gw} \propto f^{-1}$ follows from bubble collisions during the first order phase transitions~\cite{Huber:2008hg}. 

iii) The far UV zone, where the GW spectrum exhibits a clear plateau. 
While a similar feature has been present in the case of unbiased DWs, 
here it is pronounced more sharply. The IR edge of the plateau is remarkably stable: it is independent of the simulation time and the bias parameter $\epsilon$. Given that one can relate the position of the IR edge of the plateau with the scale $k/a \sim 4\pi \sqrt{H/\delta_{wall}}$, this plateau is likely to have a physical origin. 
In particular, it becomes more pronounced once the role of the artificial UV peak is reduced. Furthermore, it has been shown in Ref.~\cite{Dankovsky:2024zvs} that the plateau remains intact upon switching from the lattice $2048^3$ to $1024^3$ lattice; hence, it is unlikely to be an artefact of the limited lattice resolution at those small scales.

iv) An exponential decrease of the spectrum takes place at scales smaller than the DW width corresponding to the momenta $k/a \gtrsim 2\pi/\delta_{wall}$. This feature is independent of the potential bias, it also occurs for unbiased DWs.

\begin{table}[h]
    \centering
    \begin{tabular}{|c|c|c|ccc|}
    \hline
        $\epsilon $ &  Lattice & Reals &$a$ & $b$ & $c$ \\
        \hline
            $0$  & $2048^3$ & 1 & $2.36 \pm 0.37$ & $1.45 \pm 0.07$ & $0.74 \pm 0.02$  \\
             $0.018$  & $2048^3$ & 1 & $2.60 \pm 0.29$ & $0.97 \pm 0.01$ & $0.82 \pm 0.13$  \\
        $0.025$  & $2048^3$ & 1 & $2.29 \pm 0.14$ & $0.78 \pm 0.01$ & $0.34 \pm 0.05$ \\
       $ 0.035$  & $2048^3$ & 1 & $2.60 \pm 0.33$ & $0.92 \pm 0.02$ & $0.66 \pm 0.12$\\ 
         $0.05$  & $2048^3$ &  1 & $2.63 \pm 0.59$ & $0.86 \pm 0.04$ & $0.86 \pm 0.28$\\
    $0.05$  & $1024^3$  &   10 & $2.58 \pm 0.50$ & $1.00 \pm 0.01$ &  $0.55 \pm 0.21$\\
    \hline
    \end{tabular}
       \caption{Fitting parameters $a, b, c$ entering Eq.~\eqref{fitting} are shown for various values of the bias parameter $\epsilon$, the value $\epsilon=0$ corresponds to the unbiased case.}\label{table}
\end{table}

In the near peak region, we can also consider the following fitting formula for the GW spectrum in the peak region~\cite{NANOGrav:2023hvm}: 
\begin{equation}
\label{fitting}
\Omega_{gw} =\Omega_{gw, peak} \cdot \frac{(a+b)^c}{(bx^{-a/c}+ax^{b/c})^c} \; ,
\end{equation}
where $x \equiv f/f_{peak}$, and $a, b, c$ are the fitting parameters. We have reconstructed the latter again using \texttt{curve\_fit} function from \texttt{SciPy} Python library. The fitting procedure has been applied to GW spectra with oscillatory features suppressed, as in the case of power-law fits discussed above. Results are shown in Table~\ref{table}, where we also compare the cases $\epsilon \neq 0$ with the unbiased case, $\epsilon=0$; fit to the average over 10 simulation spectrum is performed with $\chi^2_\nu=1.63$. 
We have taken GW spectrum produced by unbiased DWs from Fig.~1 in Ref.~\cite{Dankovsky:2024zvs}; see the top panel there corresponding to vacuum initial conditions. 
We observe that introducing a bias leads to the decrease of power in the IR part of the spectrum and increase in the UV part.

\section{Conclusions} 
\label{sec:conclusions}

The slight breaking of $Z_2$-symmetry is commonly introduced  for solving the DW-domination problem. In the present work, we have numerically studied the impact of the potential bias $V_{bias}$, originating from the symmetry breaking, on the DW network evolution in a radiation-dominated Universe and the consequent production of GWs.
It appears that the common way of determining the DW collapse time through the estimate $\sigma_{wall} H_{ann} \sim V_{bias} \sim \epsilon v^3$ 
may be misleading. A more accurate estimate reads~\eqref{lawphys}, corresponding to the annihilation time~(\ref{tann}). This difference, while looking relatively innocuous, 
can lead to dramatic results in the regime of very small $\epsilon$. In particular, due to an earlier network collapse compared to naive expectations, one expects a parametrically weaker GW signal for the same bias parameter $\epsilon$. Thus, in order to have an observable GW signal, one should assume a considerably smaller potential bias $V_{bias} \propto \epsilon$ than previously thought. However, note that the $\epsilon$-dependence of the annihilation time has been inferred 
from a rather narrow range of values of $\epsilon$. In the future, owing to higher-resolution simulations and hence longer simulation times, one should be able to probe very small values of $\epsilon$ and establish the 
dependence $t_{ann} (\epsilon)$ on firm grounds.

It is worth stressing that our results regarding the DW annihilation time, while being at odds 
with naive theoretical expectations, do not contradict the numerical data obtained in the literature. Indeed, in Sec.~\ref{sec:dw} we have found that 
our results are in a comfortable agreement with the results of numerical simulations of Ref.~\cite{Pujolas}. Notably, the latter deals with particularly small $\epsilon$ values, 
for which the system reaches scaling with a high accuracy, and hence one can ignore the possible impact of initial conditions. However, in Ref.~\cite{Pujolas} the standard behaviour $t_{ann} \propto 1/\epsilon$ has been assumed. This is justified, if 
one sticks to a sufficiently narrow range of $\epsilon$, but discrepancy from the law $t_{ann} \propto 1/\epsilon^{2/3}$ becomes 
dramatic, once the extrapolation to tiny $\epsilon$ is considered.

From the viewpoint of observations, the most interesting piece of information is contained in the GW power spectrum shape. While the spectral shape of GWs produced by the network of biased walls is similar to that in the case $\epsilon=0$, there are important differences most notably in the UV part, where GWs from biased DWs exhibit larger power compared to the unbiased case. In this regard, we confirm the results obtained recently in Ref.~\cite{Cyr:2025nzf}, which uses a different code, but disagree with those in Ref.~\cite{Notari:2025kqq}, which exploits a modified CosmoLattice code. This tension is worth exploring in the future. In terms of the fitting formula~\eqref{fitting}, we have restricted values of the parameters~$a,~b$ and $c$. Note that the plateau part of the spectrum observed in Refs.~\cite{Ferreira:2023jbu, Dankovsky:2024zvs} for unbiased DWs is also present in the biased case, cf. Ref.~\cite{Kitajima}. Interestingly, in terms of conformal momentum, the position of IR edge of the plateau exhibits a remarkable stability, i.e., it remains constant as a function of time and potential bias. This may suggest that the plateau, if due to some physical mechanism rather than a simulation artefact, starts at $k/a \sim 4\pi \sqrt{H/\delta_{wall}}$. While there is currently no consensus on the nature of the plateau, our simulations in this paper and in Ref.~\cite{Dankovsky:2024zvs} indeed suggest that the plateau has a physical origin. 

 In the future, it will be interesting to analyse how biased DWs perform with respect to PTA data relatively to other GW sources. In particular, there is another type of DWs, i.e., melting DWs, capable of producing an observable GW signal~\cite{Ramazanov:2021eya, Babichev:2021uvl}. These melting DWs are described by a time-decreasing tension and they do not overclose the Universe, even in the case of an exact $Z_2$-symmetry. Spectral shapes of conventional constant tension (biased or unbiased) DWs and melting DWs are drastically different, especially in the close-to-maximum IR region\footnote{At the same time in the close-to-maximum UV range, spectral shapes of GWs sourced by melting DWs and biased DWs are very similar. As in Sec.~\ref{sec:gw}, we explain this similarity by noticing that this 
 part of the spectrum is sensitive to closed DWs (analogous to vacuum bubbles during first order phase transitions). Indeed, it has been demonstrated in Ref.~\cite{Dankovsky:2024ipq} that a large fraction of melting DW network is in the form of closed walls.}, where one has $\Omega_{gw} \propto f^{1.6}$ in the case of melting DWs~\cite{Dankovsky:2024ipq}. While the current PTA data releases are more in favor of melting DWs~\cite{Babichev:2023pbf}, the error bars are presently too large to clearly discriminate between the two sources. We hope that future PTA data releases will provide further observational support for one of these high energy physics scenarios.

\section*{Acknowledgments}
Numerical simulations have been performed on the computer cluster of the Theory Division of INR RAS and on the cluster ``Phoebe'' at CEICO, FZU. We are indebted to Josef Dvo\v r\'a\v cek for assistance with the lattice simulations on the cluster ``Phoebe''. E.B.\ acknowledges the support of ANR grant StronG
(ANR-22-CE31-0015-01). The work of I.D. and D.G. was supported by the scientific program of the National Center for Physics and Mathematics, section 5 ``Particle Physics and Cosmology'', stage 2023-2025. The work of I.D. was supported by the Foundation for the Advancement of Theoretical
Physics and Mathematics ``BASIS''. A.V. acknowledges the support from the Czech Science Foundation, GA\v{C}R, project number 24-13079S.


\end{document}